\documentclass[review]{elsarticle}
\usepackage{lineno, hyperref, lscape}
\hypersetup{%
  pdfborder = {0 0 0}
}

\usepackage{amsfonts,amsmath,amssymb,amsthm,mathrsfs,graphicx}

\newtheorem{theorem}{Theorem}[section]
\newtheorem{lemma}[theorem]{Lemma}
\newtheorem{proposition}[theorem]{Proposition}
\newtheorem{corollary}[theorem]{Corollary}

\theoremstyle{remark}
\newtheorem{remark}[theorem]{Remark}

\modulolinenumbers[5]

\journal{Computational Statistics \& Data Analysis}

\usepackage{subfigure}  
\usepackage{natbib}
\usepackage{color,soul}
\soulregister\cite7
\soulregister\citep7
\soulregister\ref7
\soulregister\eqref7
\soulregister\pageref7

\newcommand{\R}{\ensuremath{\mathbb{R}}}

\begin{document}

\begin{frontmatter}


  \title{Distributional Properties of Nearest-Site Angular Distances on the Sphere}
  %
  %
  %
  %

  \author[AddrLi]{Hongjun Li \corref{equalfirstauthor}}
  \ead{lihongjun69@bjfu.edu.cn}
  \author[AddrQiu]{Jiatong Sui\corref{equalfirstauthor}}
  \ead{jiatong.sui@hotmail.com}
  \author[AddrLi]{Shengpeng Mu}
  \ead{Shengpengmu@bjfu.edu.cn}
  \author[AddrQiu]{Xing Qiu \corref{mycorrespondingauthor}}
  \cortext[equalfirstauthor]{Hongjun Li and Jiatong Sui contributed equally to this work.}
  \cortext[mycorrespondingauthor]{Correspondence should be sent to: Xing Qiu.}
  \ead{xing_qiu@urmc.rochester.edu}

  \address[AddrLi]{College of Science, Beijing Forestry University, Beijing, China}
  \address[AddrQiu]{Department of Biostatistics and Computational Biology, University of Rochester, New York, U.S.A.}

  \begin{abstract}
Nearest-site distances arise in many applications involving spherical or directional domains, including global geospatial analysis, wireless communications, spherical clustering, and cosine-similarity-based data analysis. In this paper, we study the distributional and computational properties of $L_2$, the minimal angular great-circle distance from a uniformly distributed random point on a sphere to a set of prespecified sites on the same sphere. We first derive the cumulative distribution function (CDF) and probability density function (PDF) of $L_0$, the angular great-circle distance from a fixed vertex of a spherical triangle to a random point uniformly distributed within that triangle. We then extend these triangle-level results to convex spherical polygons and use spherical Voronoi diagrams, triangulations of Voronoi cells, and numerical integration to obtain computable distributional and moment formulas for $L_2$. In addition, we derive explicit formulas for selected moments of $\cos(L_2)$, which are relevant to cosine similarity and spherical data analysis. Extensive Monte Carlo simulations validate the proposed CDF, PDF, and moment formulas and demonstrate computational efficiency of our method relative to generic numerical integration and simulation-based alternatives.
  \end{abstract}

  \begin{keyword}
    Arc distance \sep angular great-circle distance \sep cosine similarity \sep distribution function \sep moments \sep spherical geometry \sep Voronoi diagram
  \end{keyword}

\end{frontmatter}

\section{Introduction}
\label{sec:Intro}

The nearest-site distance problem arises naturally in many applications involving spatial, geospatial, and directional data. Given a set of prespecified sites, one is often interested in identifying the nearest site to a random point and characterizing the distribution of the corresponding nearest-site distance. For example, in wireless communication, signal strength depends critically on the distance between a user's device (e.g., cell phone) and the \emph{nearest} base station. In global GIS analysis, satellite coverage, and communication systems over the Earth's surface, nearest-site distances are related to spatial coverage and service-region design. Related sphere-based proximity problems also arise in directional data analysis, spherical clustering, and cosine-similarity-based methods~\citep{dhillon2001concept,duwairi2015novel,smieja2015spherical}. In biomolecular structure modeling, similar nearest-site distances involving spherical atoms, molecular surfaces, and atomic neighborhoods have been used to characterize packing, contacts, and orientation-dependent structural relationships~\citep{poupon2004voronoi,song2022benchmark}. Therefore, understanding the distributional properties of nearest-site distances, including their cumulative distribution functions (CDFs), probability density functions (PDFs), moments, and related cosine-similarity moments, is useful for both statistical theory and computation.

In Euclidean domains, the Voronoi diagram (a.k.a. the Thiessen polygon or a Dirichlet tessellation) is a powerful tool for studying nearest-site distances. A Voronoi diagram partitions a space into regions, called Voronoi cells, according to the nearest-neighbor rule: each cell contains one seed (site), and all interior points of the cell are closer to its seed than to any other seed~\citep{augenbaum1985construction,edelsbrunner1986voronoi, aurenhammer1991voronoi}. Voronoi diagrams can be constructed efficiently~\citep{deBerg2000CompGeo} and have wide applications in forestry, chemistry, transportation, statistics, and computational geometry~\citep{persson1964distance,christofides1969expected,vaisman2012statistical,zhao2011voronoi}. They have also been used to model stochastic foam geometries~\citep{koll2014generation}, estimate Shannon entropy of multidimensional probability~\citep{lin1991divergence}, and approximate resampling-based distributions~\citep{villagran2016non}.

The Voronoi representation converts a global nearest-site problem into local problems within individual cells. In a planar wireless-communication example, base stations can be treated as seeds, and the network coverage region can be decomposed into Voronoi cells~\citep{okabe2009spatial,li2016moments}. Because each planar Voronoi cell can be decomposed into triangles sharing the seed as a common vertex, the distribution of the nearest-site distance can be studied through the simpler problem of the distance between a selected triangle vertex and a random interior point. Several research groups, including ours, have studied this problem in $\mathbb{R}^{2}$ when the random point is uniformly distributed on the triangle~\citep{aurenhammer1984optimal,Pure2015computing,ahmadi2014random,li2016moments}. In particular, we derived the closed-form CDF, PDF, and first two moments of this random distance~\citep{li2016moments}.

\begin{figure}[!h]
  \center
  \subfigure[]{  \label{fig:sphrVor}
    \includegraphics[height=1.8in]{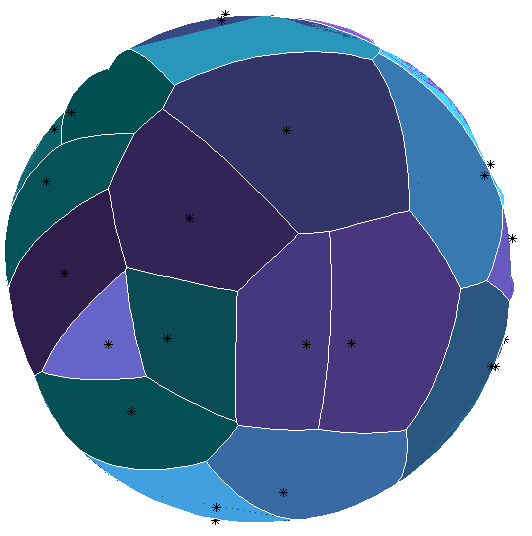} }
  \subfigure[]{  \label{fig:sphrLine}
    \includegraphics[height=1.8in]{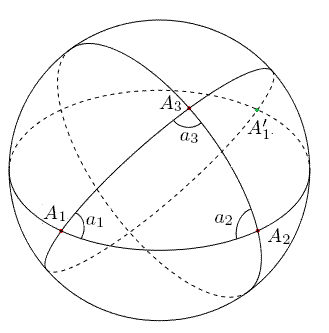} }
  \caption{An illustration of spherical Voronoi diagram and spherical triangles.
    (a) Given a set of seeds, the spherical Voronoi diagram partitions the sphere into convex spherical polygons whose points are closer, in angular great-circle distance, to the generating seed than to any other seed. The cell boundaries are portions of great-circle bisectors between pairs of seeds; triangulation is used later for computation.
    (b) Due to the non-Euclidean nature of $S^{2}$, a geodesic on $S^{2}$ is a great circle, and a spherical triangle is a region enclosed by three great-circle arcs. The spherical distance between two points such as $A_{1}$ and $A_{2}$ is measured by the length of the minor great-circle arc connecting them (obtained by multiplying the angular great-circle distance by $R$). Spherical geometry and trigonometry are fundamentally different from their planar counterparts; for instance, the sum of the interior angles of a spherical triangle exceeds 180 degrees \citep{van2012heavenly}.
  }
  \label{fig:SphrGeometry}       
\end{figure}

However, many nearest-site distance problems are not naturally planar. For global or sufficiently large spatial domains, the curvature of the Earth may become non-negligible, and planar Euclidean approximations may distort both distances and regions. In directional data analysis and spherical clustering, observations are often represented as points on a sphere, and angular great-circle distance or cosine similarity is the relevant measure of proximity. These considerations motivate the study of nearest-site distance distributions directly on spherical domains. Voronoi diagrams can be generalized to non-Euclidean geometric spaces, such as Laguerre planes and spheres~\citep{na2002voronoi,chaidee2017approximation, chaidee2018spherical}. The spherical Voronoi diagram, illustrated in Figure~\ref{fig:SphrGeometry}, is especially relevant when the underlying domain is intrinsically spherical or approximately so. For example, spherical or ellipsoidal Voronoi diagrams arise in global and large-scale geographic applications, including GIS analysis, satellite coverage, and communication systems over the Earth's surface~\citep{hu2014voronoi,liu2024dynamic}.

Despite these developments, deriving distributional properties of nearest-site distances on the sphere remains nontrivial. Spherical geometry differs fundamentally from Euclidean geometry: geodesics are great circles, distances are angular great-circle distances, and spherical triangles obey trigonometric relationships that differ from their planar counterparts. Distances between random points on $S^{2}$ have been studied in several contexts, including the generation and testing of uniformity on the sphere~\citep{bauer2000distribution}, minimization of the sum of Euclidean distances among $n$ random points~\citep{beck1984sums}, and random spherical packing~\citep{lochmann2006statistical,cai2013distributions}. These studies provide useful results for random points on spherical domains, but they do not address the nearest-site distance problem induced by a Voronoi partition on a sphere. To the best of our knowledge, no prior work has systematically studied the stochastic properties of $L_{0}(\Delta PBC)$, the angular great-circle distance from a fixed vertex $P$ of a spherical triangle to a random point $Q$ uniformly distributed on $\Delta PBC$. This triangle-level problem is the basic building block for analyzing nearest-site distances on spherical Voronoi cells.

In this study, we first derive the stochastic properties of $L_{0}(\Delta PBC)$ in detail. We then extend these results to $L_{1}(\Gamma,P)$, the corresponding distance for a convex spherical polygon $\Gamma$. Finally, using spherical Voronoi diagrams and triangulations of Voronoi cells, we derive the distribution of $L_{2}(P_{1}, \dots, P_{m})$, the minimal angular great-circle distance from a uniformly distributed random point on $S^{2}$ to a collection of prespecified sites ${P_{1}, \dots, P_{m}}$ on the same sphere. We also derive integral representations for raw moments of $L_0$, $L_1$, and $L_2$, together with explicit formulas for selected moments of $\cos(L_0)$, which are relevant to cosine similarity and spherical data analysis. The resulting framework provides a computationally tractable way to evaluate nearest-site distance distributions on spherical domains without relying solely on brute-force Monte Carlo simulation or generic numerical integration over the sphere.

This manuscript is organized as follows. Section~\ref{sec:sphereGeom} provides preliminary results on spherical triangles. Sections~\ref{sec:subTriDistr} and~\ref{sec:subPolygonDistr} derive the CDF and PDF of $L_{0}(\Delta PBC)$ and their generalization to convex spherical polygons, denoted by $L_{1}(\Gamma, P)$. Section~\ref{sec:subVoronoiDistr} derives the CDF and PDF of $L_{2}(P_{1}, \dots, P_{m})$. Section~4 provides integral representations for arbitrary raw moments of $L_0(\triangle PBC)$, $L_1(\Gamma,P)$, and $L_2(P_1,\ldots,P_m)$, which are evaluated numerically using Gaussian quadrature. \ref{sec:append-cosine-similarity} gives explicit formulas for several moments of $\cos(L_0(\triangle PBC))$, which are useful in spherical cluster analysis and data mining~\citep{dhillon2001concept,qiu2015new,nguyen2010cosine,ye2011cosine}.


\section{Useful geometric properties of spherical triangles}
\label{sec:sphereGeom}

\begin{figure}[!h]
  \center
  \subfigure[]{  \label{fig:SphrCap}
    \includegraphics[height=1.35in]{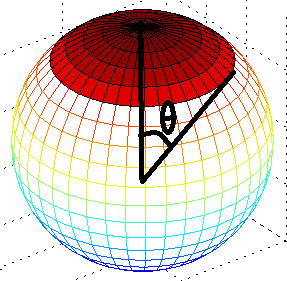} }\quad
  \subfigure[]{  \label{fig:Sphrfan}
    \includegraphics[height=1.35in]{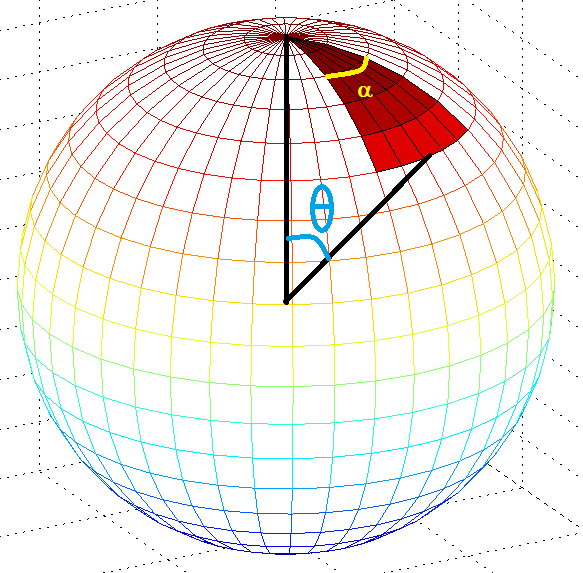} }\quad
  \subfigure[]{  \label{fig:SphrTri}
    \includegraphics[height=1.35in]{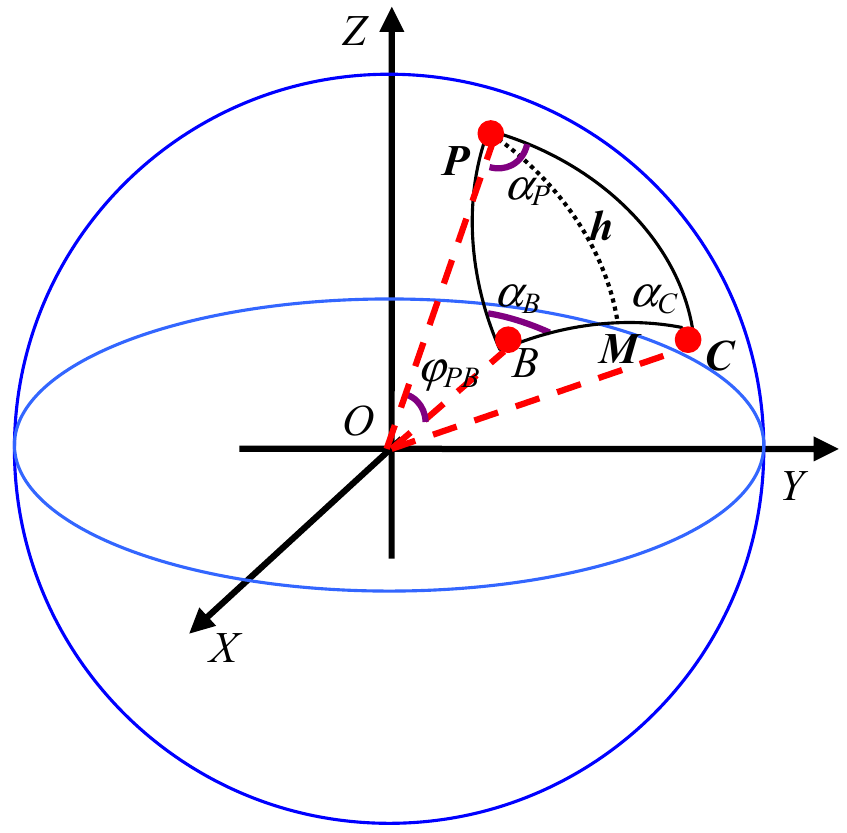} }
  \caption{The three sub-figures illustrate the cap, fan and triangle on a unit sphere.
    (a) A spherical cap is the region of a sphere cut off by a plane. If the plane passes through the center of the sphere, the cap is called a hemisphere. We could use a second plane to cut the cap, which will result in a spherical segment.
    (b) A spherical fan on the sphere is part of a spherical cap generated by further partitions, using great circles passing through the apex or pole of the spherical cap.
    (c) A spherical triangle can be formed by connecting three points on a sphere with great circle arcs. In this paper, we consider the inner triangles instead of the outer triangles. The sum of the interior angles of a spherical triangle lies between $\pi$ and $3\pi$.
  }
  \label{fig:capAndFan}       
\end{figure}

Throughout this manuscript, we denote the sphere in $\mathbb{R}^3$ centered
at the origin, denoted by $O$, with radius $R$ by
\[
S_R^2 := \{x\in\mathbb{R}^3:\|x\|=R\}.
\]
Let $A$ and $B$ be two distinct non-antipodal points on $S_R^2$, i.e.,
$A\neq \pm B$. These points define a great circle, i.e., a geodesic on $S_R^2$.
We denote by
\[
L_{AB}:=\phi_{AB}:=\arccos\left(\frac{A\cdot B}{R^2}\right)\in[0,\pi]
\]
the angular great-circle distance between $A$ and $B$, measured in radians.
The corresponding physical arc length is $R L_{AB}$. On the unit sphere, the angular great-circle distance is numerically identical
to the usual arc distance. On a sphere of radius $R$, however, the physical arc
length and angular distance differ by a factor of $R$: if $L$ denotes angular
distance in radians, then the corresponding physical arc length is $D=RL$.
Conversely, an arc length $D$ corresponds to angular distance $D/R$. Throughout the main
derivations, all distance variables, including $L_{AB}$, $L_0$, $L_1$, $L_2$,
$r$, $h$, $L_{PB}$, and $L_{PC}$, denote angular distances in radians unless
explicitly stated otherwise; physical arc-length quantities can be recovered afterward by this conversion.

Under this angular-distance convention, trigonometric expressions such as
$\sin r$, $\cos r$, $\tan r$, and $\cos(L_2)$ are dimensionally well defined.
Surface areas retain the factor $R^2$ because they are measured on the sphere
of radius $R$.

Unless otherwise stated, all spherical triangles and polygons considered below
are assumed to be geodesically convex, with edges given by minor great-circle
arcs. We also exclude degenerate configurations: no relevant pair of vertices is
antipodal, no spherical triangle has zero area, and the spherical Voronoi diagram
is assumed to be in general position. In particular, sites are distinct, no
Voronoi cell has zero area, and distance ties occur only on cell boundaries,
which have surface area zero. For example, a spherical cap with angular radius $r$ has area
$2\pi R^2(1-\cos r)$, and a spherical sector with opening angle $\alpha$ and
angular radius $r$ has area $\alpha R^2(1-\cos r)$. If physical arc-length
distances are desired, they can be recovered by multiplying angular distances
by $R$.

We also denote the area of the spherical cap by $A_{cap}$ and the half-aperture angle \citep{ahmadi2014random} of this spherical cap by $\theta$ (see Fig. \ref{fig:SphrCap}). The area of a spherical sector, shown in Fig. \ref{fig:Sphrfan}, is denoted by $A_{fan}$, and its top angle is denoted by $\alpha$. Formal definitions of these concepts can be found in~\ref{sec:append-def}.

\paragraph{Area of spherical triangle}

Let $\alpha_P, \alpha_B, \alpha_C$ be the three interior angles of triangle $\Delta PBC$, as shown in Fig. \ref{fig:SphrTri}. It is well known that the area of $\Delta PBC$ is
\begin{equation}
  \label{eq:SphrTriArea}
  A_{\Delta} = R^2 (\alpha_P + \alpha_B + \alpha_C - \pi )=R^2 E
\end{equation}

Here $\alpha_P, \alpha_B$, and $\alpha_C$ represent angles between great circles. The spherical angle $\angle APB = \alpha$ can be calculated with the following Equation
\begin{equation}
  \label{eq:SphrArcAngle}
  \alpha = \arccos \big( ( \overrightarrow{OA} \times \overrightarrow{OP})^0 \cdot (\overrightarrow{OB}\times \overrightarrow{OP})^0 \big)
\end{equation}
Here $\overrightarrow{v}^{0}$ represents the unit direction of vector $v \in \R^{3}$, namely, $\overrightarrow{v}^{0} := \overrightarrow{v} / |\overrightarrow{v}|$.

The calculation for $\alpha_B$ and $\alpha_C$ is similar.
$E$ is called the \textit{spherical excess}, $0 \leq E \leq 2 \pi$, with $E=0$ being the degenerate case in which all three points lie on one great circle.

\paragraph{Altitude of spherical triangle}
An important concept used in our derivation is
the angular altitude of a spherical triangle, which is a generalization of the
altitude of a planar triangle~\citep{li2016moments}. For the spherical triangle
$\triangle PBC$, the angular altitude to the base arc $\widehat{BC}$ is the
central angle $\angle POM$, as shown in Fig.~2(c), and is denoted by $h_{BC}$
or simply $h$. It can be calculated as
\begin{equation}
h = \angle POM = \arcsin\left(\left|\vec n\cdot \overrightarrow{OP}^{\,0}\right|\right),
\end{equation}
where $\vec n$ is the unit normal direction of the plane defined by points
$O$, $B$, and $C$:
\[
\vec n := (\overrightarrow{OB}\times \overrightarrow{OC})^0.
\]
Thus $h$ is the shortest angular distance from vertex $P$ to the great circle
defined by $B$ and $C$. The corresponding physical altitude length is $Rh$.

In addition, the foot point $M$ on the great circle determined by $B$ and $C$
can be computed by projecting the unit direction of $P$ onto the plane
orthogonal to $\vec n$ and then normalizing back to the sphere. Let
\[
p:=\frac{\overrightarrow{OP}}{R},\qquad
u:=p-(p\cdot \vec n)\vec n .
\]
Provided $u\neq 0$, the foot point is
\begin{equation}
  \label{eq:AltitudePointPosi}
  \overrightarrow{OM}=R\frac{u}{\|u\|}.
\end{equation}
This equation is used only to determine the foot point $M$ geometrically. It
does not change the convention that $h$ and all subsequent distance variables
are angular distances. The degenerate case $u=0$ corresponds to $P$ being a
pole of the great circle determined by $B$ and $C$ and is excluded from the
generic configurations considered below.

\section{Distribution of the Minimal Angular Great-Circle Distance}
\label{sec:distanceDistr}

In this section, we derive the distribution of $L_2(P_1,\ldots,P_m)$, the
minimal angular great-circle distance from a random point $Q$ to a prespecified
set of sites $\{P_1,\ldots,P_m\}$ on a sphere.

\subsection{Distribution of $L_0(\triangle PBC)$}
\label{sec:subTriDistr}
Let $\triangle PBC$ be an arbitrary spherical triangle on $S_R^2$, and let $Q$ be a random point uniformly distributed on $\triangle PBC$. Define
\[
L_0 \equiv L_0(\triangle PBC) := L_{PQ},
\]
the angular great-circle distance from the fixed vertex $P$ to the random
point $Q$.

\begin{figure}[!h]
  \center
  \subfigure[]{  \label{fig:AltitudIn}
    \includegraphics[height=1.45in]{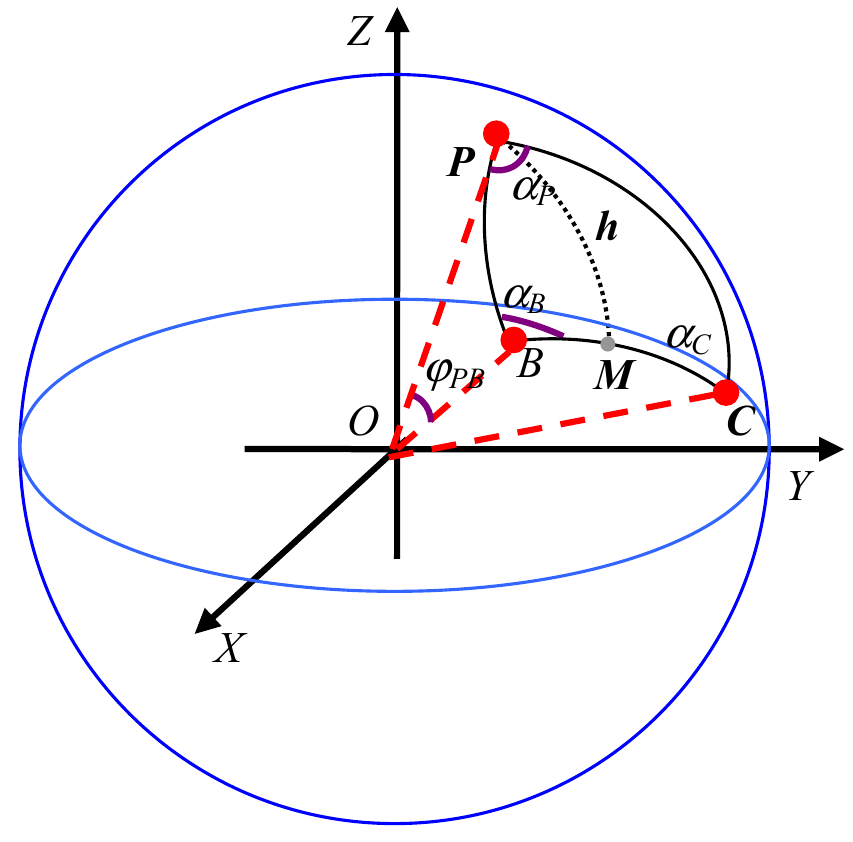} }
  \subfigure[]{  \label{fig:AltitudOut}
    \includegraphics[height=1.45in]{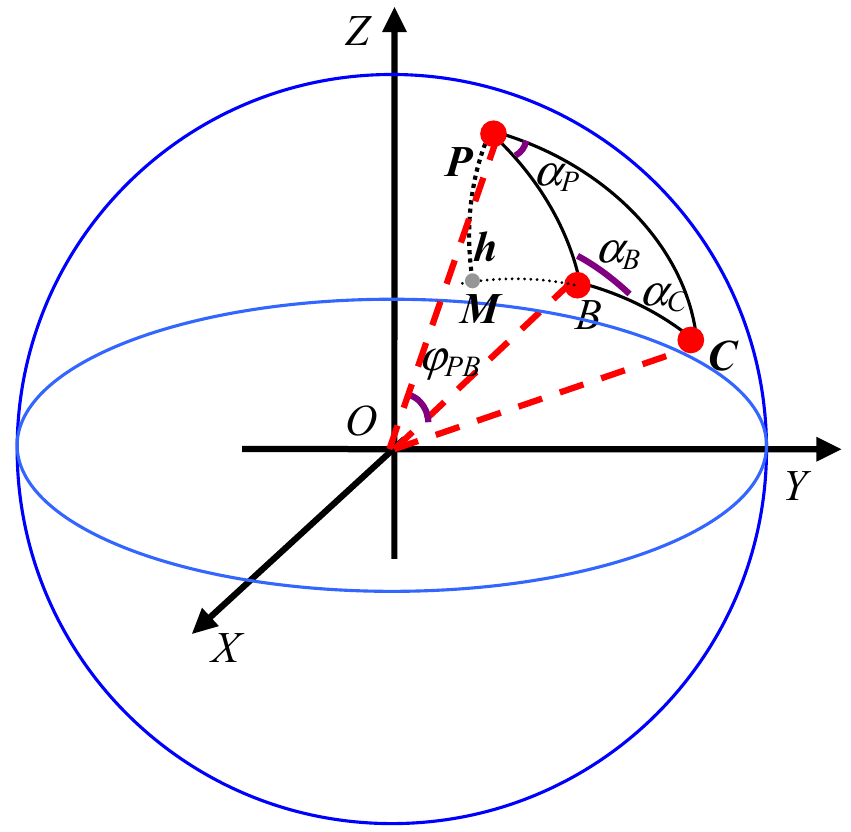} }
  \subfigure[]{  \label{fig:AltitudMerge}
    \includegraphics[height=1.45in]{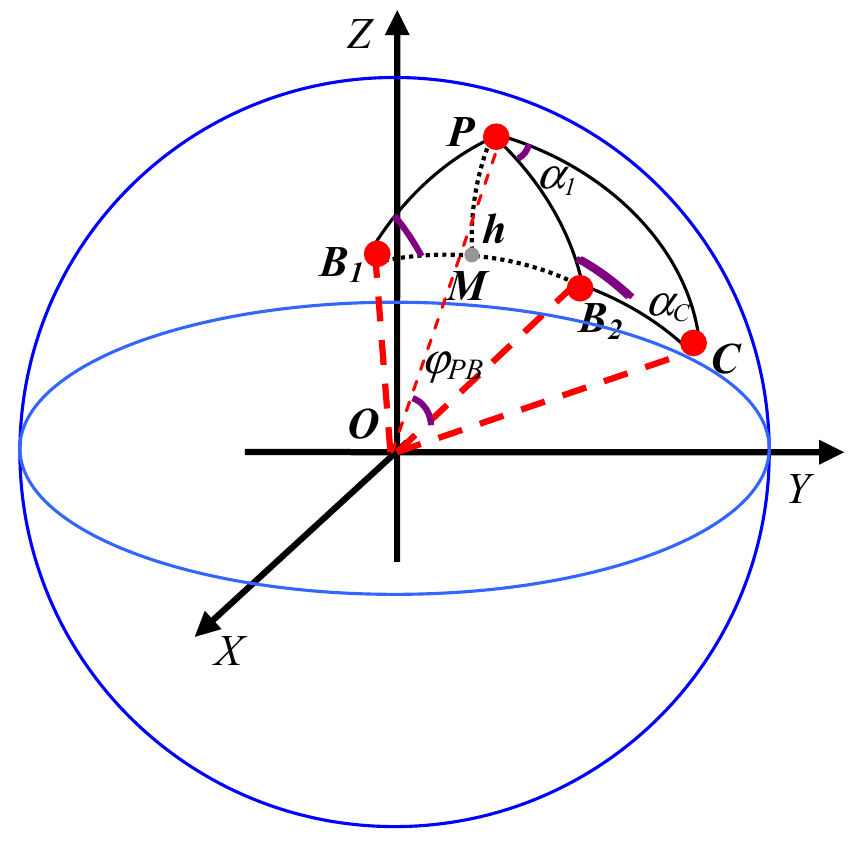} }
  \caption{$\Delta PBC$ can be classified into two classes based on the location of the altitude $PM$:
    (a) The inside-altitude case, $PM \subset \Delta PBC$,
    (b) the outside-altitude case, $PM\not\subset \Delta PBC$ (equivalently, the foot point $M$ lies outside the base arc $\widehat{BC}$).
    By reflecting arc $PB$ about the altitude $PM$, we can unify these two cases, which is illustrated in (c).  }
  \label{fig:AltitudesandMerge}       
\end{figure}

Let
\[
\Omega:=|\triangle PBC|
\]
denote the area of $\triangle PBC$. Without loss of generality, assume
\[
L_{PB}\le L_{PC}.
\]
In the formulas below, $h$ denotes the first angular radius at which the cap
centered at $P$ meets the base side $\widehat{BC}$ under the reflected-arc
construction described below. Equivalently, we restrict attention to the generic
geometric configurations for which the relevant cap-intersection thresholds can
be ordered as
\[
0\le h\le L_{PB}\le L_{PC}.
\]
If a triangle falls outside this ordering, the same cap-intersection principle
still applies, but the breakpoints must be reordered and the corresponding
piecewise branches adjusted.
Let $h$ be the altitude from $P$ to the base arc $\widehat{BC}$, and let $M$ be the foot of that altitude, as defined in Section 2. As noted by Ahmadi and Pan~\cite{ahmadi2014random}, spherical triangles may be classified according to whether the altitude from $P$ lies inside or outside the triangle. In the present work, these two configurations are unified by the reflected-arc construction shown in Figure~\ref{fig:AltitudIn} and Figure~\ref{fig:AltitudOut}: reflect the arc $\widehat{PB}$ about the altitude $PM$, and denote the reflected arcs by $\widehat{PB_1}$ and $\widehat{PB_2}$. In the inside-altitude case, $B=B_1$; in the outside-altitude case, $B=B_2$. This construction allows both geometric configurations to be handled within a single analytic framework.

For $r\ge 0$, let $B_r(P)$ denote the spherical cap centered at $P$ with
angular radius $r$, and define
\[
S(r):=\bigl|B_r(P)\cap \triangle PBC\bigr|.
\]
By construction, the event $\{L_0\le r\}$ is exactly the set of points in $\triangle PBC$ lying within angular great-circle distance $r$ from $P$. Hence
\[
F_{L_0}(r)=\mathbb{P}(L_0\le r)=\frac{S(r)}{\Omega}.
\]
Therefore, the distribution of $L_0$ reduces to the evaluation of the intersection area $S(r)$. The geometry of this intersection depends on how the cap boundary meets the sides of $\triangle PBC$, which leads to the following four regimes.

\begin{lemma}
Under the ordered-threshold condition $0\le h\le L_{PB}\le L_{PC}$, as $r$
increases from $0$, the intersection $B_r(P)\cap \triangle PBC$ passes through
four geometric regimes:
\begin{enumerate}
    \item $0<r\le h$: the intersection is a spherical sector centered at $P$;
    \item $h<r\le L_{PB}$: the intersection consists of one spherical triangle and two spherical sectors;
    \item $L_{PB}<r\le L_{PC}$: the intersection consists of one spherical triangle and one spherical sector;
    \item $r>L_{PC}$: the cap contains the entire triangle $\triangle PBC$.
\end{enumerate}
\end{lemma}

\begin{figure}[!h]
  \center
  \subfigure[]{  \label{fig:triFanCase01}
    \includegraphics[height=2.2in]{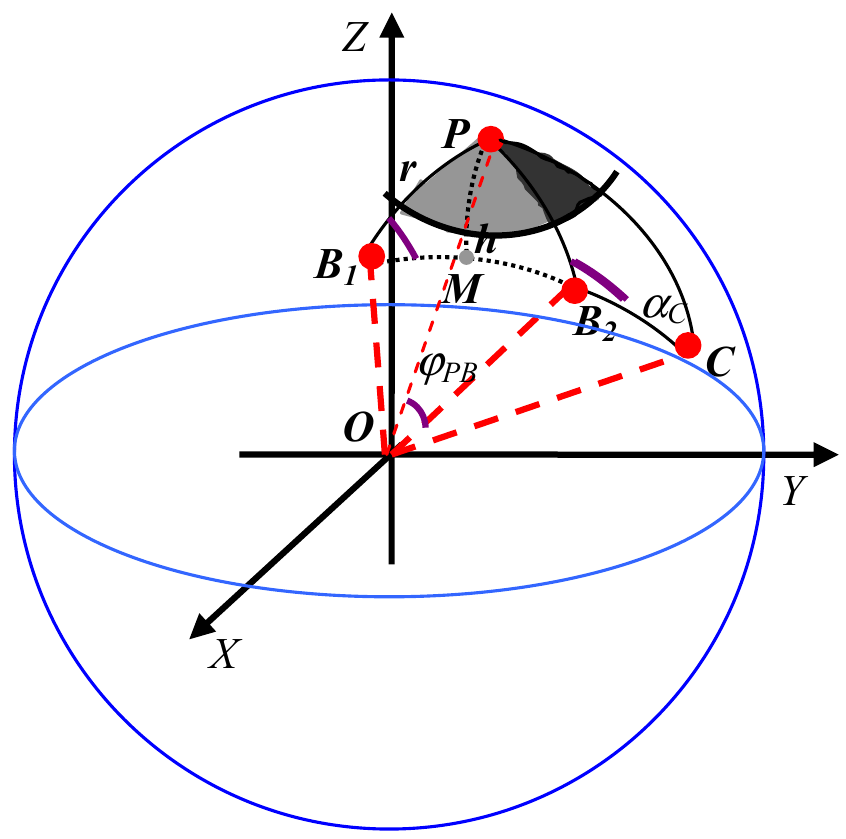} }
  \subfigure[]{  \label{fig:triFanCase02}
    \includegraphics[height=2.2in]{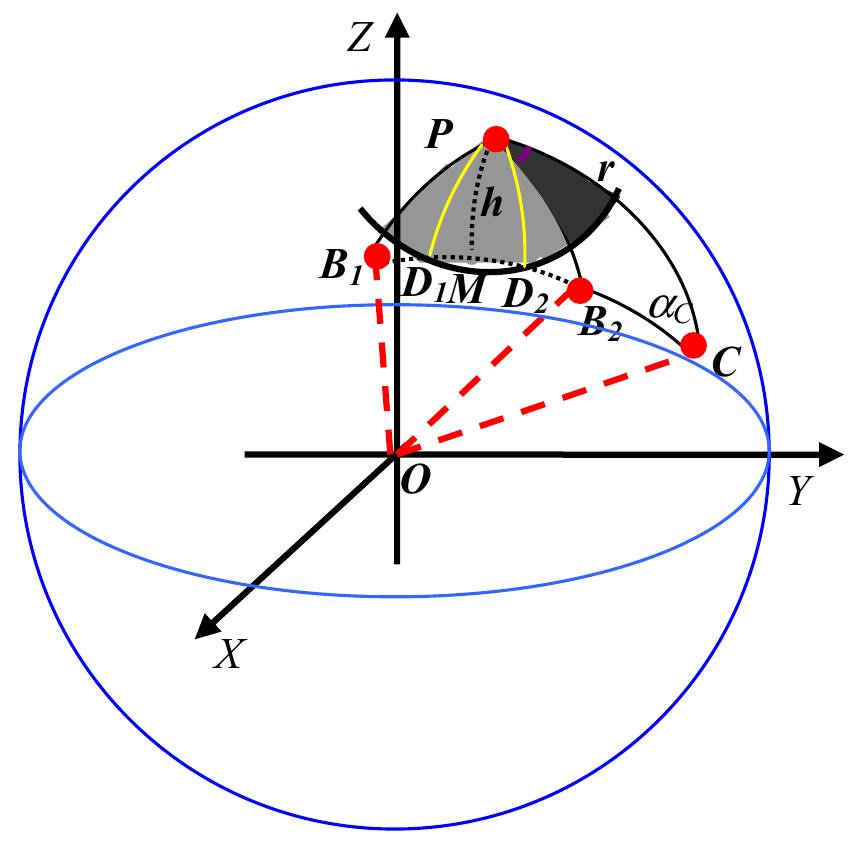} } \\
  \subfigure[]{  \label{fig:triFanCase03}
    \includegraphics[height=2.2in]{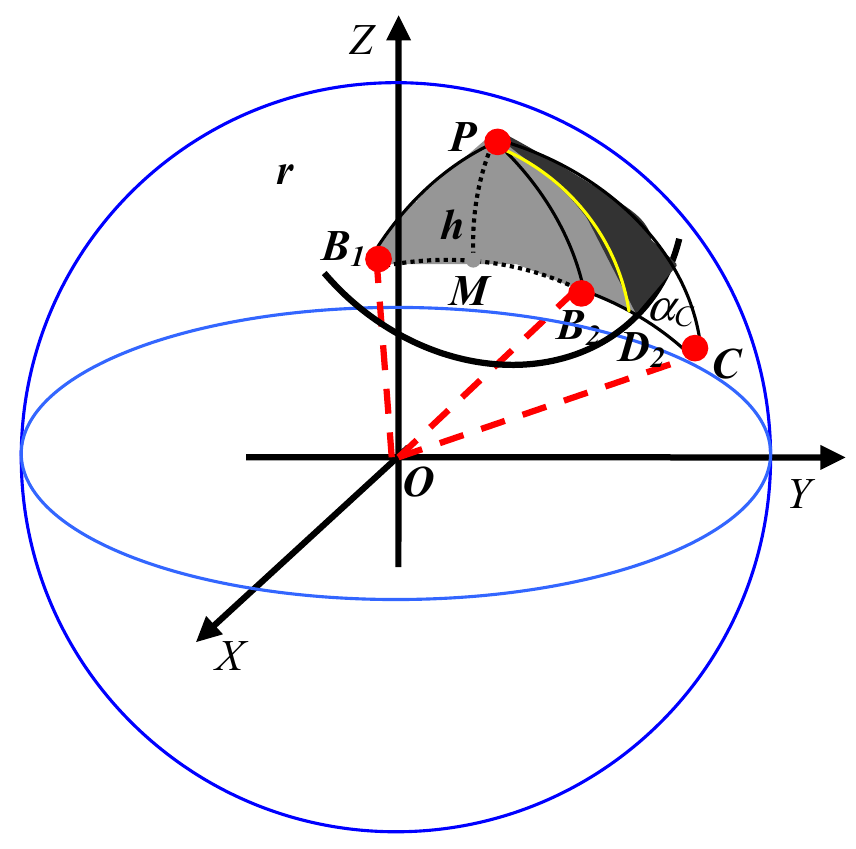} }
  \subfigure[]{  \label{fig:triFanCase04}
    \includegraphics[height=2.2in]{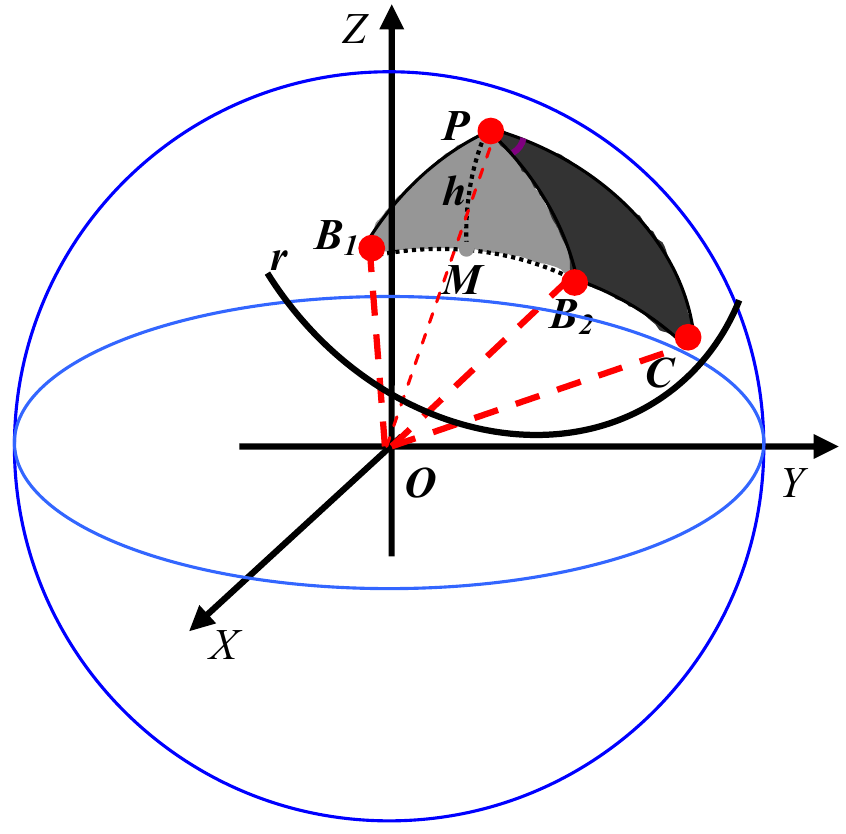} }
  \caption{The CDF/PDF of $L_{0}(\Delta PBC)$ depends on how the spherical fan originated from $P$ with angular radius $r$ intersects with $\Delta PBC$. By construction, every point $Q$ in this fan satisfies $L_{PQ} \le r$. Therefore, $P(L_{0} \leqslant r)$ equals the ratio of the intersection area to the area of the triangle. This intersection can be classified as follows:
    (a) case 1: $0 < r \le h$, the intersection is a sector.
    (b) case 2: $h < r \le L_{PB}$, it contains a spherical triangle and two sectors.
    (c) case 3: $L_{PB} < r \le L_{PC}$, it contains a spherical triangle and a sector.
    (d) case 4: $ r > L_{PC}$, it is a spherical triangle. }
  \label{fig:triSctCases}       
\end{figure}

\begin{proof}
Under the ordered-threshold condition, the cap boundary first reaches the base
side $\widehat{BC}$ at angular radius $h$, then the endpoint on the shorter side
from $P$ at $L_{PB}$, and finally the endpoint on the longer side from $P$ at
$L_{PC}$. These events determine the four possible forms of the cap--triangle
intersection. The corresponding configurations are illustrated in
Figure~\ref{fig:triSctCases}.
\end{proof}

To express the area formulas compactly, we introduce several auxiliary angles.

\begin{lemma}
For $h<r\le L_{PC}$, define
\[
\alpha_r:=\angle MPD_2,
\qquad
\beta_r:=\angle PD_2M.
\]
Then
\begin{equation}
\alpha_r
=
\cos^{-1}
\!\left[
(\overrightarrow{OM}\times \overrightarrow{OP})_0
\cdot
(\overrightarrow{OD_2}\times \overrightarrow{OP})_0
\right]
=
\arccos\!\left(\frac{\tan h}{\tan r}\right),
\label{eq:alpha_r}
\end{equation}
and
\begin{equation}
\beta_r
=
\cos^{-1}
\!\left[
(\overrightarrow{OP}\times \overrightarrow{OD_2})_0
\cdot
(\overrightarrow{OM}\times \overrightarrow{OD_2})_0
\right]
=
\arcsin\!\left(\frac{\sin h}{\sin r}\right).
\label{eq:beta_r}
\end{equation}

For $L_{PB}<r\le L_{PC}$, define
\[
\omega_r:=\angle D_2PC.
\]
Then
\begin{equation}
\omega_r
=
\cos^{-1}
\!\left[
(\overrightarrow{OC}\times \overrightarrow{OP})_0
\cdot
(\overrightarrow{OD_2}\times \overrightarrow{OP})_0
\right]
=
\angle MPC-\alpha_r.
\label{eq:omega_r}
\end{equation}
\end{lemma}

\begin{proof}
These identities follow from the spherical right-triangle geometry induced by the altitude $PM$, the cap boundary of radius $r$, and the intersection point $D_2$. The reflected-arc construction makes the same formulas valid for both the inside-altitude and outside-altitude cases.
\end{proof}

We now derive the area of the cap--triangle intersection.

\begin{proposition}
Let $S(r)=|B_r(P)\cap \triangle PBC|$. Then the following formulas hold.

\textbf{Case 1: $0<r\le h$.}
In this range, $B_r(P)\cap \triangle PBC$ is a spherical sector with opening angle $\alpha_P=\angle BPC$. Hence
\begin{equation}
S(r)=\alpha_P(1-\cos r)R^2.
\label{eq:S_case1}
\end{equation}

\textbf{Case 2: $h<r\le L_{PB}$.}
In this range, $B_r(P)\cap \triangle PBC$ consists of the spherical triangle $\triangle PD_1D_2$ together with two spherical sectors. Let
\[
\alpha_1:=\angle CPB_2,
\qquad
\alpha_2:=\angle MPB_2,
\]
and define
\[
\eta:=\frac{L_{BB_2}}{L_{MB_2}},
\]
so that $\eta=0$ if $B=B_2$ and $\eta=2$ if $B=B_1$. Then
\begin{equation}
S(r)=
\left\{
\alpha_1(1-\cos r)
+
\eta\left[
\left(\frac{\pi}{2}+\alpha_r+\beta_r-\pi\right)
+
(\alpha_2-\alpha_r)(1-\cos r)
\right]
\right\}R^2.
\label{eq:S_case2}
\end{equation}

\textbf{Case 3: $L_{PB}<r\le L_{PC}$.}
In this range, $B_r(P)\cap \triangle PBC$ consists of the spherical triangle $\triangle PBD_2$ and one spherical sector. Let $\alpha_B=\angle PBC$. Then
\begin{equation}
S(r)=
\left[
(\alpha_P-\omega_r+\alpha_B+\beta_r-\pi)
+
\omega_r(1-\cos r)
\right]R^2.
\label{eq:S_case3}
\end{equation}

\textbf{Case 4: $r>L_{PC}$.}
In this range, the cap contains the entire spherical triangle, and therefore
\begin{equation}
S(r)=\Omega.
\label{eq:S_case4}
\end{equation}
\end{proposition}

\begin{proof}
For $0<r\le h$, the cap intersects the triangle in a single sector centered at $P$, so Equation~\eqref{eq:S_case1} follows directly from the spherical-sector area formula.

For $h<r\le L_{PB}$, the intersection is the union of $\triangle PD_1D_2$ and two sectors. Applying the spherical triangle area formula together with the sector area formula yields Equation~\eqref{eq:S_case2}. The factor $\eta$ encodes the two possible altitude configurations in a unified way.

For $L_{PB}<r\le L_{PC}$, the intersection is the union of $\triangle PBD_2$ and one sector, which gives Equation~\eqref{eq:S_case3}.

For $r>L_{PC}$, the cap covers all of $\triangle PBC$, so $S(r)=\Omega$.
\end{proof}

The previous proposition can be summarized in a single piecewise formula.

\begin{theorem}\label{thm:S_piecewise}
Let
\[
g_r:=1-\cos r.
\]
Then, for all $r\in\mathbb{R}$,
\begin{equation}
S(r)=
\begin{cases}
0, & r\le 0,\\[4pt]
\alpha_P g_r R^2, & 0<r\le h,\\[4pt]
\left(\alpha_1 g_r+\eta\left[(\alpha_r+\beta_r-\pi/2)+(\alpha_2-\alpha_r)g_r\right]\right)R^2,
& h<r\le L_{PB},\\[6pt]
\left((\alpha_P+\alpha_B+\beta_r-\pi)-\omega_r\cos r\right)R^2,
& L_{PB}<r\le L_{PC},\\[6pt]
\Omega, & r>L_{PC}.
\end{cases}
\label{eq:S_piecewise}
\end{equation}
\end{theorem}

\begin{proof}
This follows by combining the four cases in Proposition 3.3 and noting that
\[
(\alpha_P-\omega_r+\alpha_B+\beta_r-\pi)+\omega_r(1-\cos r)
=
(\alpha_P+\alpha_B+\beta_r-\pi)-\omega_r\cos r.
\]
The branch $r\le 0$ is included only to extend the CDF to all real numbers.
\end{proof}

\begin{corollary}
The cumulative distribution function of $L_0(\triangle PBC)$ is
\begin{equation}
F_{L_0}(r)=\frac{S(r)}{\Omega},
\label{eq:CDF_L0}
\end{equation}
where $S(r)$ is given by Theorem~\ref{thm:S_piecewise}.
\end{corollary}

\begin{proof}
Because $Q$ is uniformly distributed on $\triangle PBC$,
\[
\mathbb{P}(L_0\le r)
=
\frac{|B_r(P)\cap \triangle PBC|}{|\triangle PBC|}
=
\frac{S(r)}{\Omega}.
\]
\end{proof}

To obtain the density, we differentiate the piecewise formula for $S(r)$. The required derivatives of the auxiliary angles are stated next.

\begin{lemma}
For $h\le r\le L_{PC}$,
\begin{equation}
\alpha_r'
=
\frac{d\alpha_r}{dr}
=
\frac{\tan h}{\sin r\cos r\sqrt{\tan^2 r-\tan^2 h}},
\label{eq:alpha_prime}
\end{equation}
\begin{equation}
\beta_r'
=
\frac{d\beta_r}{dr}
=
-\frac{\sin h\cos r}{\sin r\sqrt{\sin^2 r-\sin^2 h}},
\label{eq:beta_prime}
\end{equation}
and
\begin{equation}
\omega_r'=-\alpha_r'.
\label{eq:omega_prime}
\end{equation}
\end{lemma}

\begin{proof}
Differentiate the expressions for $\alpha_r$ and $\beta_r$ in Equations~\eqref{eq:alpha_r} and \eqref{eq:beta_r} by the chain rule. Since $\omega_r=\angle MPC-\alpha_r$, and $\angle MPC$ is constant with respect to $r$, it follows immediately that $\omega_r'=-\alpha_r'$.
\end{proof}

\begin{corollary}
\label{cor:PDF_L0}
The probability density function of $L_0(\triangle PBC)$ is
\[
f_{L_0}(r)=\frac{S'(r)}{\Omega},
\]
that is,
\begin{equation}
f_{L_0}(r)=
\begin{cases}
0, & r\le 0 \text{ or } r>L_{PC},\\[6pt]
\dfrac{\alpha_P\sin r\,R^2}{\Omega}, & 0<r\le h,\\[10pt]
\dfrac{\left\{\alpha_1\sin r+\eta\left[\alpha_r'\cos r+\beta_r'+(\alpha_2-\alpha_r)\sin r\right]\right\}R^2}{\Omega},
& h<r\le L_{PB},\\[12pt]
\dfrac{\left[\beta_r'-\omega_r'\cos r+\omega_r\sin r\right]R^2}{\Omega},
& L_{PB}<r\le L_{PC}.
\end{cases}
\label{eq:PDF_L0}
\end{equation}
\end{corollary}

\begin{proof}
Differentiate Theorem~\ref{thm:S_piecewise} branch by branch.

For $0<r\le h$,
\[
S(r)=\alpha_P(1-\cos r)R^2
\quad\Rightarrow\quad
S'(r)=\alpha_P\sin r\,R^2.
\]

For $h<r\le L_{PB}$,
\[
S(r)=
\left(\alpha_1 g_r+\eta\left[(\alpha_r+\beta_r-\pi/2)+(\alpha_2-\alpha_r)g_r\right]\right)R^2,
\]
hence
\[
S'(r)=
\left\{
\alpha_1\sin r
+
\eta\left[\alpha_r'\cos r+\beta_r'+(\alpha_2-\alpha_r)\sin r\right]
\right\}R^2.
\]

For $L_{PB}<r\le L_{PC}$,
\[
S(r)=
\left((\alpha_P+\alpha_B+\beta_r-\pi)-\omega_r\cos r\right)R^2,
\]
so
\[
S'(r)=
\left[\beta_r'-\omega_r'\cos r+\omega_r\sin r\right]R^2.
\]

Outside $(0,L_{PC}]$, the function $S(r)$ is constant, so the density is zero. Dividing by $\Omega$ completes the proof.
\end{proof}

\begin{remark}
The formulas above are understood branchwise on the intervals $(0,h)$,
$(h,L_{PB})$, and $(L_{PB},L_{PC})$. The CDF remains continuous at the
breakpoints because the area function $S(r)$ is defined geometrically and the
adjacent formulas represent the same cap--triangle intersection at the boundary
values. The density is therefore interpreted in the usual almost-everywhere
sense; values at the finitely many breakpoints $h$, $L_{PB}$, and $L_{PC}$ may
be assigned arbitrarily without changing the distribution. In numerical
implementation, the endpoint values should be evaluated by one-sided limits or
avoided in quadrature rules because the separate auxiliary derivatives
$\alpha'_r$ and $\beta'_r$ may be singular at $r=h$ even though the
cap-intersection area itself is continuous.
\end{remark}

\begin{remark}
The reflected-arc construction is the key device that allows the inside-altitude and outside-altitude configurations to be handled by a single unified derivation. This is why the same formulas for $S(r)$, $F_{L_0}(r)$, and $f_{L_0}(r)$ apply in both cases.
\end{remark}

\subsection{Extension to Convex Spherical Polygons}
\label{sec:subPolygonDistr}

In many applications, one is interested in the distribution of the angular great-circle distance from a prespecified interior point of a spherical polygon to a random point uniformly distributed in that polygon. In this subsection, we extend the results of Section~3.1 from a spherical triangle to a convex spherical polygon.

Let $\Gamma \subset S_R^2$ be a convex spherical polygon with vertices $B_1,B_2,\dots,B_n$, listed in cyclic order, and let $P$ be a prespecified interior point of $\Gamma$. Let $Q$ be uniformly distributed on $\Gamma$, and define
\[
L_1 \equiv L_1(\Gamma,P):=L_{PQ},
\]
the angular great-circle distance from $P$ to $Q$.

\begin{figure}[!ht]
  \center
  \includegraphics[height=3.5in]{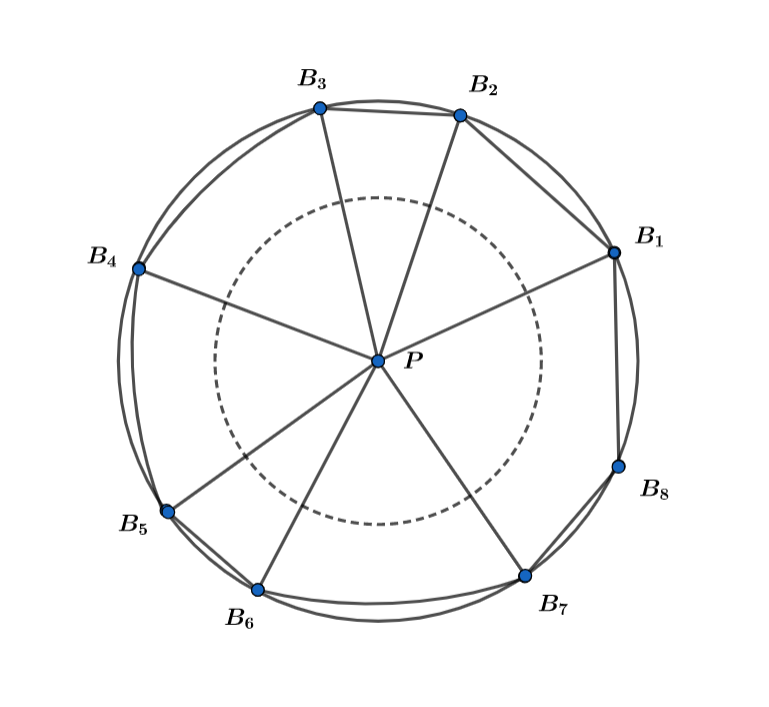}
  \caption{A convex spherical polygon $\Gamma$ can be decomposed into a set of adjacent spherical triangles sharing a prespecified interior point $P$ as a common vertex. In Fig.\ref{fig:sphrPolyandR}, the spherical cap centered at $P$ represents a subset of $\Gamma$ in which every point $Q$ satisfies $L_{PQ} \leqslant r$, where $r$ is the angular radius of this spherical cap. Apparently, this spherical cap is the disjoint union of spherical fans pertaining to the component triangles. Consequently, the statistical properties of $L_{1} = L_{PQ}$, the angular great-circle distance between $P$ and a random point $Q$ in $\Gamma$, can be derived from the properties of $L_0$, the random angular great-circle distance from a random point $Q$ to a selected vertex $P$ of a spherical triangle.  
  \label{fig:sphrPolyandR}}
\end{figure}

Because $\Gamma$ is convex and $P$ lies in its interior, the polygon can be decomposed into $n$ spherical triangles sharing $P$ as a common vertex:
\[
\Gamma=\bigcup_{i=1}^{n}\triangle PB_iB_{i+1},
\qquad B_{n+1}:=B_1,
\]
where the union is disjoint up to boundaries. This decomposition is illustrated in Figure~\ref{fig:sphrPolyandR}. For each $i=1,\dots,n$, define
\[
S_i(r):=\bigl|B_r(P)\cap \triangle PB_iB_{i+1}\bigr|.
\]
Then the distribution of $L_1$ follows by summing the triangle-level contributions obtained in Section~3.1.

\begin{proposition}
\label{prop:L1_distribution}
Let $\Gamma\subset S_R^2$ be a convex spherical polygon with vertices $B_1,\dots,B_n$, and let $P$ be an interior point of $\Gamma$. Let $Q$ be uniformly distributed on $\Gamma$, and define
\[
L_1(\Gamma,P):=L_{PQ}.
\]
Then the cumulative distribution function and probability density function of $L_1(\Gamma,P)$ are given by
\begin{equation}
F_{L_1}(r)
=
\mathbb{P}(L_{PQ}\le r)
=
\frac{1}{|\Gamma|}
\sum_{i=1}^{n}S_i(r),
\qquad
f_{L_1}(r)
=
\frac{1}{|\Gamma|}
\sum_{i=1}^{n}S_i'(r),
\label{eq:L1_cdf_pdf}
\end{equation}
where
\[
S_i(r)=\bigl|B_r(P)\cap \triangle PB_iB_{i+1}\bigr|,
\qquad i=1,\dots,n,
\]
with $B_{n+1}:=B_1$. Moreover, each function $S_i(r)$ and its derivative $S_i'(r)$ can be computed using the formulas in Theorem~\ref{thm:S_piecewise} and Corollary~\ref{cor:PDF_L0}, applied to the spherical triangle $\triangle PB_iB_{i+1}$.
\end{proposition}

\begin{proof}
Since $Q$ is uniformly distributed on $\Gamma$,
\[
\mathbb{P}(L_{PQ}\le r)
=
\frac{\bigl|\{Q\in\Gamma:L_{PQ}\le r\}\bigr|}{|\Gamma|}
=
\frac{|B_r(P)\cap \Gamma|}{|\Gamma|}.
\]
Because $\Gamma$ is convex and $P$ lies in its interior, it can be triangulated into the spherical triangles $\triangle PB_iB_{i+1}$, $i=1,\dots,n$, which are disjoint up to boundaries. Therefore,
\[
|B_r(P)\cap \Gamma|
=
\sum_{i=1}^{n}|B_r(P)\cap \triangle PB_iB_{i+1}|
=
\sum_{i=1}^{n}S_i(r),
\]
and hence
\[
F_{L_1}(r)=\frac{1}{|\Gamma|}\sum_{i=1}^{n}S_i(r).
\]
Differentiating branchwise with respect to $r$ yields
\[
f_{L_1}(r)=\frac{1}{|\Gamma|}\sum_{i=1}^{n}S_i'(r).
\]
Finally, since each $\triangle PB_iB_{i+1}$ is a spherical triangle with vertex $P$, the formulas derived in Section~3.1 apply to each $S_i(r)$ and $S_i'(r)$ separately. Summing these contributions over $i=1,\dots,n$ completes the proof.
\end{proof}

\begin{remark}
Equation~\eqref{eq:L1_cdf_pdf} shows that the distribution of $L_1(\Gamma,P)$ is a weighted aggregation of triangle-level distributions. Indeed,
\[
S_i(r)
=
|\triangle PB_iB_{i+1}|
\cdot
\mathbb{P}(L_{PQ}\le r \mid Q\in \triangle PB_iB_{i+1}),
\]
so that
\[
F_{L_1}(r)
=
\sum_{i=1}^{n}
\frac{|\triangle PB_iB_{i+1}|}{|\Gamma|}
\,
\mathbb{P}(L_{PQ}\le r \mid Q\in \triangle PB_iB_{i+1}).
\]
Thus $F_{L_1}(r)$ may be interpreted as a weighted sum of conditional CDFs over the component triangles.
\end{remark}

\subsection{Distribution of the Minimal Angular Great-circle Distance $L_2(P_1,\dots,P_m)$}
\label{sec:subVoronoiDistr}

\begin{figure}[!ht]
  \center
  \includegraphics[height=3.5in]{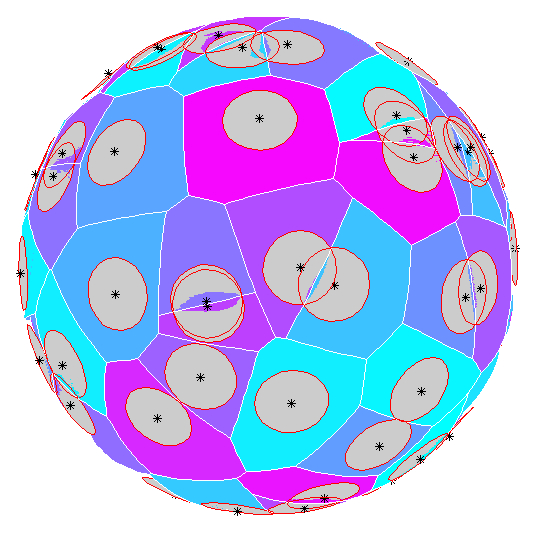}
  \caption{A spherical Voronoi diagram is a partition of the sphere that consists of adjacent and disjoint spherical polygons ($\left\{ \Gamma_{1}, \dots, \Gamma_{m} \right\}$) based on prespecified seeds $\left\{ P_{1}, \dots, P_{m} \right\}$. For each point $Q \in \Gamma_{j}$, its angular great-circle distance to $P_{j}$ (the seed of $\Gamma_{j}$) is the shortest among all seeds. Circles in the figure represent points that have the same arc distance to a given seed. The intersections of these circles can be used to determine the boundaries between adjacent cells.
  }
  \label{fig:sphrVoronoiandR}       
\end{figure}
Let $P_1,P_2,\dots,P_m$ be a prespecified set of points on $S_R^2$, and let $Q$ be a random point uniformly distributed on the sphere. Define
\[
L_2 := \min_{1\le j\le m} L_{P_jQ},
\]
the minimal angular great-circle distance from $Q$ to the set
$\{P_1,P_2,\ldots,P_m\}$.

To derive the distribution of $L_2$, we partition the sphere into the spherical Voronoi cells generated by the sites $P_1,\dots,P_m$. Let
\[
\Gamma_1,\Gamma_2,\dots,\Gamma_m
\]
denote this spherical Voronoi diagram. By construction, each cell $\Gamma_j$ is a spherical polygon such that every point in $\Gamma_j$ is closer to $P_j$ than to any other seed in terms of angular great-circle distance. Therefore, if $Q\in \Gamma_j$, then
\[
L_2=L_{P_jQ}.
\]
This observation reduces the distribution of $L_2$ to a weighted combination of the conditional distributions within the individual Voronoi cells, as illustrated in Figure~\ref{fig:sphrVoronoiandR}.

For each $j=1,\dots,m$, triangulate $\Gamma_j$ into spherical triangles
\[
\triangle P_jB_{i,j}B_{i+1,j},
\qquad i=1,\dots,n_j,
\]
where $n_j$ is the number of vertices of $\Gamma_j$. For each such triangle, define
\[
S_{i,j}(r):=
\left|B_r(P_j)\cap \triangle P_jB_{i,j}B_{i+1,j}\right|.
\]
Then the distribution of $L_2$ follows immediately from the polygon-level result in Section~3.2.

\begin{theorem}\label{thm:L2_distribution}
Let $P_1,\dots,P_m\in S_R^2$ be prespecified sites, let $Q$ be uniformly distributed on $S_R^2$, and define
\[
L_2:=\min_{1\le j\le m}L_{P_jQ}.
\]
Let $\Gamma_1,\dots,\Gamma_m$ denote the spherical Voronoi cells generated by $P_1,\dots,P_m$. For each $j$, triangulate $\Gamma_j$ into spherical triangles
\[
\triangle P_jB_{i,j}B_{i+1,j},
\qquad i=1,\dots,n_j,
\]
and define
\[
S_{i,j}(r):=
\left|B_r(P_j)\cap \triangle P_jB_{i,j}B_{i+1,j}\right|.
\]
Then the cumulative distribution function and probability density function of $L_2$ are given by
\begin{equation}
F_{L_2}(r)
=
\sum_{j=1}^{m}\frac{|\Gamma_j|}{4\pi R^2}\,F_{L_1}(r\mid Q\in \Gamma_j)
=
\frac{1}{4\pi R^2}\sum_{j=1}^{m}\sum_{i=1}^{n_j}S_{i,j}(r),
\label{eq:L2_cdf}
\end{equation}
and
\begin{equation}
f_{L_2}(r)
=
F_{L_2}'(r)
=
\frac{1}{4\pi R^2}\sum_{j=1}^{m}\sum_{i=1}^{n_j}S_{i,j}'(r).
\label{eq:L2_pdf}
\end{equation}
\end{theorem}

\begin{proof}
Since $Q$ is uniformly distributed on the sphere, the probability that $Q$ lies in $\Gamma_j$ is
\[
\mathbb{P}(Q\in \Gamma_j)=\frac{|\Gamma_j|}{4\pi R^2}.
\]
By the defining property of the spherical Voronoi diagram, every point in $\Gamma_j$ is closest to $P_j$. Therefore, conditional on $Q\in \Gamma_j$,
\[
L_2=L_{P_jQ}.
\]
It follows that
\[
F_{L_2}(r)
=
\sum_{j=1}^{m}
\mathbb{P}(Q\in \Gamma_j)\,
\mathbb{P}(L_2\le r\mid Q\in \Gamma_j)
=
\sum_{j=1}^{m}
\frac{|\Gamma_j|}{4\pi R^2}\,
F_{L_1}(r\mid Q\in \Gamma_j).
\]

Now apply Proposition~\ref{prop:L1_distribution} to each Voronoi cell $\Gamma_j$. Since $\Gamma_j$ is triangulated into
\[
\triangle P_jB_{1,j}B_{2,j},\dots,\triangle P_jB_{n_j,j}B_{1,j},
\]
we have
\[
F_{L_1}(r\mid Q\in \Gamma_j)
=
\frac{1}{|\Gamma_j|}
\sum_{i=1}^{n_j}S_{i,j}(r).
\]
Substituting this into the previous expression gives
\[
F_{L_2}(r)
=
\frac{1}{4\pi R^2}
\sum_{j=1}^{m}\sum_{i=1}^{n_j}S_{i,j}(r).
\]
Differentiating branchwise with respect to $r$ yields
\[
f_{L_2}(r)
=
\frac{1}{4\pi R^2}
\sum_{j=1}^{m}\sum_{i=1}^{n_j}S_{i,j}'(r),
\]
which proves the result.
\end{proof}

\begin{remark}
Theorem~\ref{thm:L2_distribution} shows that the distribution of $L_2$ is obtained by averaging the polygon-level distributions over the Voronoi partition of the sphere. Equivalently, $L_2$ is a mixture distribution whose components are the conditional distances $L_{P_jQ}$ within the cells $\Gamma_j$, with weights proportional to the cell areas:
\[
\frac{|\Gamma_j|}{4\pi R^2},
\qquad j=1,\dots,m.
\]
Thus the nearest-site problem on the sphere reduces to a finite sum of triangle-level cap-intersection problems already solved in Section~3.1 and aggregated in Section~3.2.
\end{remark}

\section{Moments}
\label{sec:Moments}

With the density functions derived in Section~3, we now derive integral
representations for the raw moments of $L_0$, $L_1$, and $L_2$. These formulas
apply to arbitrary positive integer orders and can be evaluated numerically
using standard quadrature methods. Recall that $L_0$ denotes the angular
great-circle distance from a fixed vertex $P$ of a spherical triangle
$\triangle PBC$ to a uniformly distributed random point $Q$ in that triangle,
$L_1(\Gamma,P)$ denotes the corresponding angular distance for a convex
spherical polygon $\Gamma$ with prespecified interior point $P$, and $L_2$
denotes the minimal angular great-circle distance from a uniformly distributed
random point on the sphere to a set of prespecified sites $\{P_1,\ldots,P_m\}$.

\begin{proposition}
\label{prop:moment_L0}
Let $L_0=L_0(\triangle PBC)$, and let $f_{L_0}(r)$ be the density given in Corollary~\ref{cor:PDF_L0}. Then, for each integer $k\ge 1$,
\begin{equation}
\mathbb{E}\!\left(L_0^k(\triangle PBC)\right)
=
\int_{-\infty}^{\infty} r^k f_{L_0}(r)\,dr
=
\frac{R^2}{\Omega}\bigl(\mathrm{Term}_1+\mathrm{Term}_2+\mathrm{Term}_3\bigr),
\label{eq:moment_L0}
\end{equation}
where
\[
\mathrm{Term}_1
=
\int_0^h r^k\,\alpha_P\sin r\,dr,
\]
\[
\mathrm{Term}_2
=
\int_h^{L_{PB}}
r^k\Bigl(\alpha_1\sin r+\eta\bigl(\alpha_r'\cos r+\beta_r'+(\alpha_2-\alpha_r)\sin r\bigr)\Bigr)\,dr,
\]
and
\[
\mathrm{Term}_3
=
\int_{L_{PB}}^{L_{PC}}
r^k\bigl(\beta_r'-\omega_r'\cos r+\omega_r\sin r\bigr)\,dr.
\]
\end{proposition}

\begin{proof}
By definition,
\[
\mathbb{E}(L_0^k)=\int_{-\infty}^{\infty} r^k f_{L_0}(r)\,dr.
\]
Since the density $f_{L_0}(r)$ is piecewise defined on the intervals $(0,h]$, $(h,L_{PB}]$, and $(L_{PB},L_{PC}]$, we substitute the corresponding branches of Corollary~\ref{cor:PDF_L0} and split the integral accordingly. This yields the three-term representation above.
\end{proof}

\begin{remark}
The functions $\alpha_r'$ and $\beta_r'$ appearing in Proposition~\ref{prop:moment_L0} are given by Equations~\eqref{eq:alpha_prime} and \eqref{eq:beta_prime}. To the best of our knowledge, the integral representation in Equation~\eqref{eq:moment_L0} does not appear to admit a tractable closed-form solution in general. In practice, it can be evaluated numerically using a suitable integration scheme; in our implementation, Gaussian quadrature is used.
\end{remark}

The polygon-level and Voronoi-level moments follow immediately from the weighted decomposition results in Sections~3.2 and 3.3.

\begin{corollary}
\label{cor:moment_L1}
Let $\Gamma\subset S_R^2$ be a convex spherical polygon, let $P$ be a prespecified interior point of $\Gamma$, and let $L_1=L_1(\Gamma,P)$. If $\Gamma$ is decomposed into spherical triangles
\[
\triangle PB_1B_2,\dots,\triangle PB_nB_1,
\]
then
\begin{equation}
\mathbb{E}\!\left(L_1^k(\Gamma)\right)
=
\frac{1}{|\Gamma|}
\sum_{i=1}^{n}
|\triangle PB_iB_{i+1}|\,
\mathbb{E}\!\left(L_0^k(\triangle PB_iB_{i+1})\right).
\label{eq:moment_L1}
\end{equation}
\end{corollary}

\begin{proof}
Condition on the component triangle containing $Q$. Since $Q$ is uniformly distributed on $\Gamma$, the probability that $Q$ lies in $\triangle PB_iB_{i+1}$ is
\[
\frac{|\triangle PB_iB_{i+1}|}{|\Gamma|}.
\]
The conditional moment on that event is exactly $\mathbb{E}(L_0^k(\triangle PB_iB_{i+1}))$. Summing over all component triangles yields the result.
\end{proof}

\begin{corollary}
Let $P_1,\dots,P_m\in S_R^2$ be prespecified sites, let $Q$ be uniformly distributed on the sphere, and let
\[
L_2:=\min_{1\le j\le m}L_{P_jQ}.
\]
Let $\Gamma_1,\dots,\Gamma_m$ denote the corresponding spherical Voronoi cells. Then
\[
\mathbb{E}\!\left(L_2^k(\{P_1,\dots,P_m\})\right)
=
\frac{1}{4\pi R^2}
\sum_{j=1}^{m}
|\Gamma_j|\,\mathbb{E}(L_1^k(\Gamma_j)),
\]
or equivalently,
\begin{equation}
\mathbb{E}\!\left(L_2^k(\{P_1,\dots,P_m\})\right)
=
\frac{1}{4\pi R^2}
\sum_{j=1}^{m}
\sum_{i=1}^{n_j}
|\triangle P_jB_{i,j}B_{i+1,j}|\,
\mathbb{E}\!\left(L_0^k(\triangle P_jB_{i,j}B_{i+1,j})\right).
\label{eq:moment_L2}
\end{equation}
\end{corollary}

\begin{proof}
Condition on the Voronoi cell containing $Q$. Since $Q$ is uniformly distributed on the sphere, the probability that $Q\in \Gamma_j$ is
\[
\frac{|\Gamma_j|}{4\pi R^2}.
\]
Conditional on $Q\in \Gamma_j$, the distance $L_2$ coincides with the polygon-level distance from $P_j$ to $Q$, whose $k$th moment is $\mathbb{E}(L_1^k(\Gamma_j))$. Averaging over Voronoi cells gives the first formula. The second follows by applying Corollary~\ref{cor:moment_L1} to each $\Gamma_j$.
\end{proof}

\begin{remark}
The moments in Proposition~4.1 and Corollaries~4.3--4.4 are moments
of angular distances. If $D_a=R L_a$ denotes the corresponding physical
arc-length distance for $a\in\{0,1,2\}$, then
\[
E(D_a^k)=R^k E(L_a^k).
\]
Similarly, the CDF and PDF of $D_a$ are obtained from those of $L_a$ by
\[
F_{D_a}(d)=F_{L_a}(d/R),\qquad
f_{D_a}(d)=\frac{1}{R}f_{L_a}(d/R).
\]
\end{remark}

Finally, the same strategy can be used to derive moments of cosine similarity.
Because $L_a$ is an angular great-circle distance, $\cos(L_a)$ is the normalized
inner product between the corresponding direction vectors on the sphere. In
particular, by substituting the density of $L_0$ into integrals of the form
\[
E\{\cos^k(L_a)\}, \qquad a\in\{0,1,2\}, \quad k\in\{1,2,4,6\},
\]
one obtains explicit formulas for several cosine moments. The triangle-level
formulas for $\cos^k(L_0)$ are derived in Appendix~B. The corresponding
polygon-level and Voronoi-level cosine moments for $L_1$ and $L_2$ are then
obtained by the same weighted aggregation used in Corollaries~4.3 and~4.4.

\section{Numerical Experiments}

We conducted numerical experiments to validate the theoretical results for
the minimal angular great-circle distance
\[
L_2=\min_{1\le j\le m} L_{P_jQ},
\]
where $Q$ is uniformly distributed on $S_R^2$ and $P_1,\ldots,P_m$ are
prespecified sites on the sphere. The experiments serve two complementary purposes. First, we validate the full \emph{CDF} and \emph{PDF} derived in Section~3.3 by direct comparison with Monte Carlo simulations. Second, for continuity with the earlier version of the manuscript, we also assess the accuracy of the theoretical formulas for several moments of $L_2$ and $\cos(L_2)$.

Throughout, we set $m=100$ and consider the same three site-generation mechanisms as in the original numerical study.

\textbf{Sim1.} All 100 sites are independently and uniformly sampled on $S_R^2$.

\textbf{Sim2.} The first 75 sites are sampled uniformly on the sphere and reflected, if necessary, into the upper hemisphere, while the remaining 25 sites are sampled uniformly and reflected, if necessary, into the lower hemisphere.

\textbf{Sim3.} 50 sites are sampled uniformly \citep{marsaglia1972choosing} from the northern cap $\{(x,y,z)\in S_R^2:z\ge 0.9\}$, and the remaining 50 are sampled uniformly from the southern cap $\{(x,y,z)\in S_R^2:z\le -0.9\}$.

\begin{figure}[!h]
  \center
  \subfigure[Sim1]{  \label{fig:Sim1}
    \includegraphics[height=1.19in]{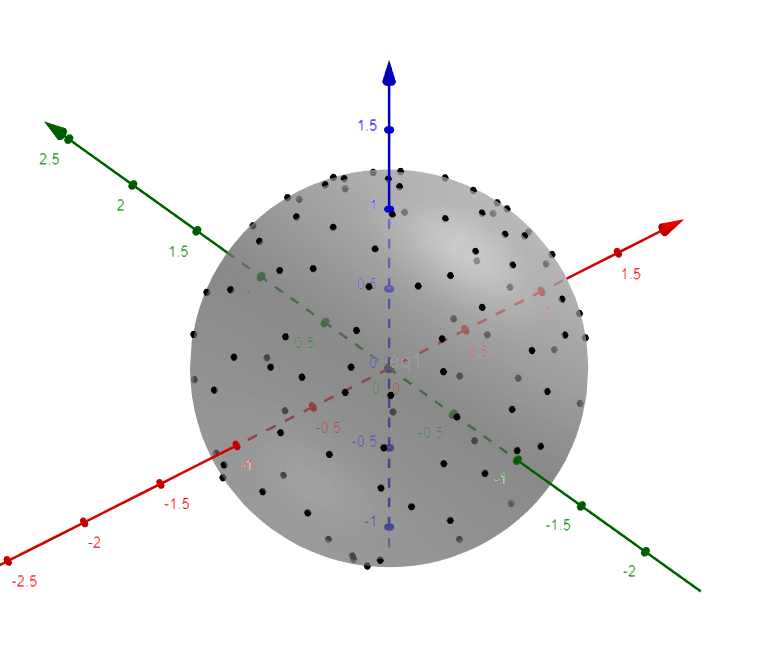} }
  \subfigure[Sim2]{  \label{fig:Sim2}
    \includegraphics[height=1.19in]{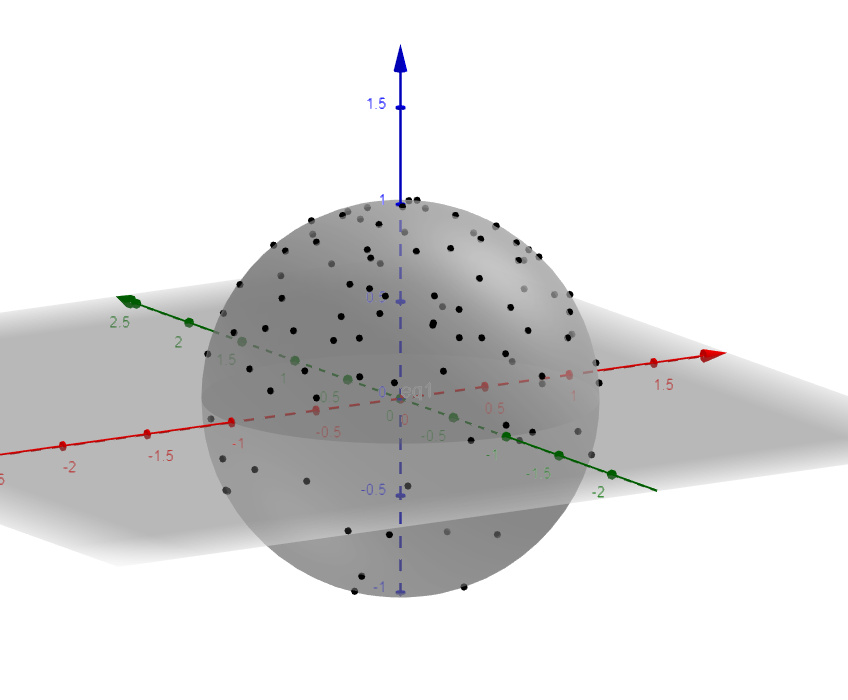} }
  \subfigure[Sim3]{  \label{fig:Sim3}
    \includegraphics[height=1.19in]{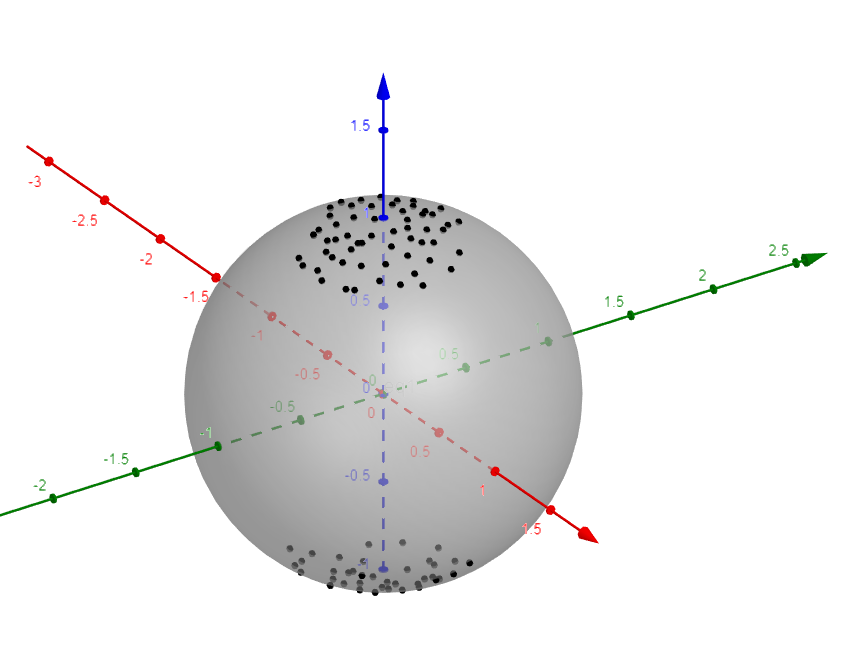} }
  \caption{Three distributions of seeds used in simulations.  (a) In Sim1, all $m=100$ seeds are uniformly generated on the surface of the sphere.
    (b) In Sim2, the majority of the seeds (75 of them) are uniformly distributed on the upper hemisphere of the sphere, while the rest (25 of them) are uniformly distributed on the lower hemisphere.
    (c) In Sim3, half of the sites are uniformly generated on the North end region, while others are uniformly generated on the South end region.
  }
  \label{fig:Sim_Fig}       
\end{figure}

For each scenario, one site configuration is first generated and then held fixed. Conditional on this fixed seed configuration, we perform 100 independent Monte Carlo repetitions, each based on 10,000 uniformly sampled query points on the sphere. This fixed-seed design ensures that the empirical results are compared against the same theoretical distribution associated with the same Voronoi tessellation.

\subsection{Distributional validation of the CDF and PDF}

We first validate the full distributional formulas for $L_2$. For each fixed site configuration, we compute the theoretical CDF and PDF from the Voronoi-based expression in Section~3.3. In each Monte Carlo repetition, we generate 10,000 query points on the sphere, compute the corresponding nearest-site distances, and then construct an empirical CDF and an empirical histogram.

\begin{figure}[htbp]
\centering

\begin{minipage}{0.24\textwidth}
    \centering
    \includegraphics[width=\textwidth]{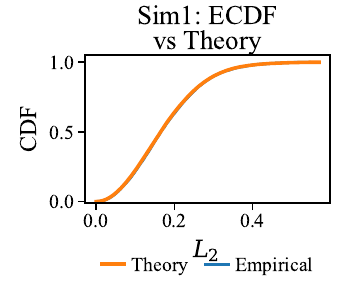}
    
\end{minipage}
\hfill
\begin{minipage}{0.24\textwidth}
    \centering
    \includegraphics[width=\textwidth]{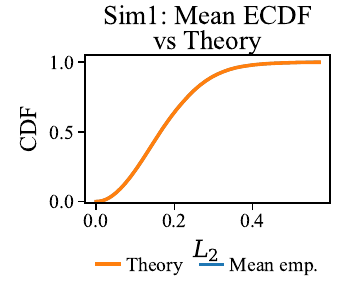}
    
\end{minipage}
\hfill
\begin{minipage}{0.24\textwidth}
    \centering
    \includegraphics[width=\textwidth]{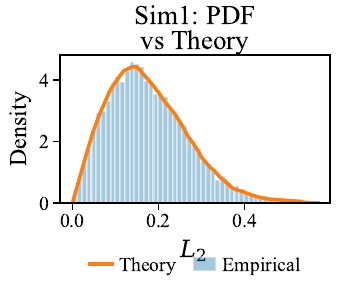}
    
\end{minipage}
\hfill
\begin{minipage}{0.24\textwidth}
    \centering
    \includegraphics[width=\textwidth]{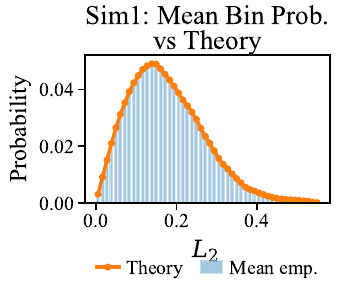}
    
\end{minipage}

\vspace{0.6em}

\begin{minipage}{0.24\textwidth}
    \centering
    \includegraphics[width=\textwidth]{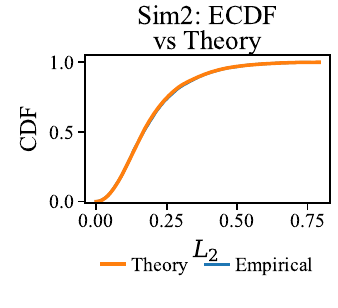}
    
\end{minipage}
\hfill
\begin{minipage}{0.24\textwidth}
    \centering
    \includegraphics[width=\textwidth]{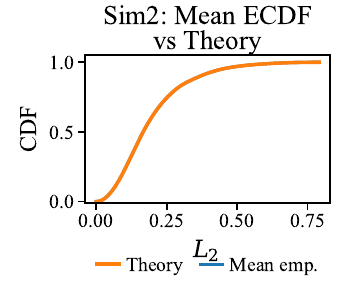}
    
\end{minipage}
\hfill
\begin{minipage}{0.24\textwidth}
    \centering
    \includegraphics[width=\textwidth]{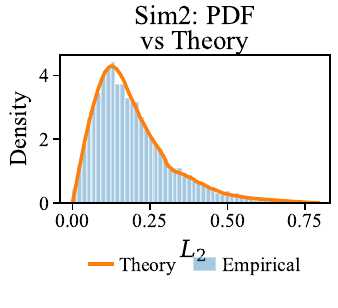}
    
\end{minipage}
\hfill
\begin{minipage}{0.24\textwidth}
    \centering
    \includegraphics[width=\textwidth]{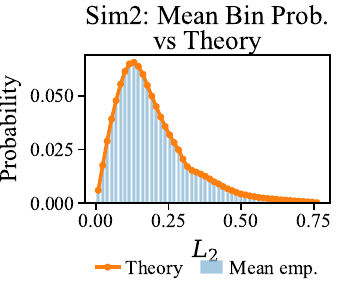}
    
\end{minipage}

\vspace{0.6em}

\begin{minipage}{0.24\textwidth}
    \centering
    \includegraphics[width=\textwidth]{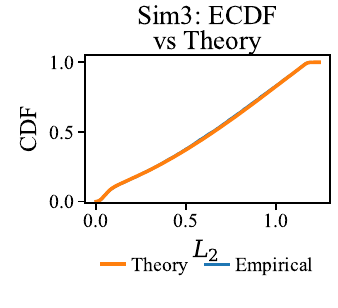}
    
\end{minipage}
\hfill
\begin{minipage}{0.24\textwidth}
    \centering
    \includegraphics[width=\textwidth]{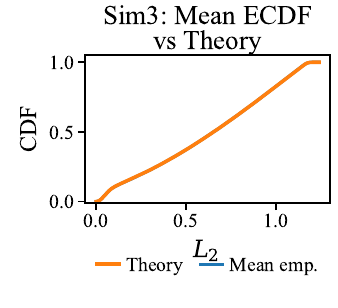}
    
\end{minipage}
\hfill
\begin{minipage}{0.24\textwidth}
    \centering
    \includegraphics[width=\textwidth]{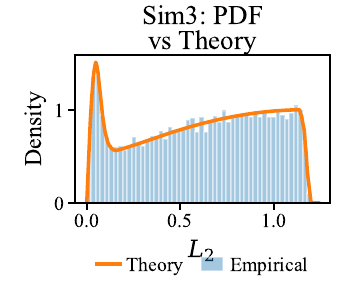}
    
\end{minipage}
\hfill
\begin{minipage}{0.24\textwidth}
    \centering
    \includegraphics[width=\textwidth]{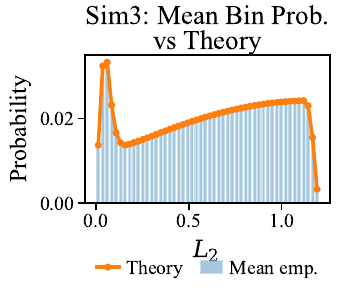}
    
\end{minipage}

\caption{Numerical validation of the CDF and PDF of $L_2$ under Sim1--Sim3. Rows correspond to Sim1, Sim2, and Sim3. Columns correspond to: representative empirical step CDF versus theoretical CDF; mean empirical CDF over 100 repetitions versus theoretical CDF; representative empirical histogram versus theoretical PDF; and mean empirical bin probabilities over 100 repetitions versus theoretical bin probabilities.}
\label{fig:cdf_pdf_validation}
\end{figure}

For visualization, Figure~\ref{fig:cdf_pdf_validation} reports, for each scenario, four complementary plots:
(a) one representative empirical \emph{step CDF} together with the theoretical CDF;
(b) the \emph{mean empirical CDF} over the 100 repetitions together with the theoretical CDF;
(c) one representative empirical \emph{histogram} together with the theoretical PDF; and
(d) the \emph{mean empirical bin probabilities} over the 100 repetitions together with the corresponding theoretical bin probabilities.
The rows correspond to Sim1, Sim2, and Sim3, respectively.

To quantify the discrepancy between empirical and theoretical distributions, we report the following metrics. For the CDF, we use a Kolmogorov-type discrepancy,
\[
\mathrm{KS}=\sup_r |\widehat F_n(r)-F(r)|,
\]
together with the integrated absolute and squared errors,
\[
\mathrm{CDF}\text{-}L^1=\int |\widehat F_n(r)-F(r)|\,dr,
\qquad
\mathrm{CDF}\text{-}L^2=\int (\widehat F_n(r)-F(r))^2\,dr.
\]
For the PDF, rather than comparing raw histogram heights directly, we compare empirical and theoretical \emph{bin probabilities}. If $0=t_0<t_1<\cdots<t_K$ denotes a bin partition, the empirical and theoretical probabilities in bin $k$ are denoted by $\widehat p_k$ and $p_k=F(t_k)-F(t_{k-1})$, respectively, and we summarize the discrepancy by
\[
\mathrm{PDF\_bin\_IAE}=\sum_{k=1}^K |\widehat p_k-p_k|,
\qquad
\mathrm{PDF\_bin\_ISE}=\sum_{k=1}^K (\widehat p_k-p_k)^2.
\]

Figure~\ref{fig:cdf_pdf_validation} shows close agreement between simulation and theory in all three scenarios. In Sim1 and Sim2, the empirical step CDFs are nearly indistinguishable from the theoretical CDFs, and the mean empirical CDFs over 100 repetitions lie essentially on top of the theoretical curves. On the PDF side, the representative empirical histograms track the theoretical PDFs closely, while the mean empirical bin probabilities also agree well with the corresponding theoretical bin probabilities. In Sim3, where the sites are concentrated in two polar caps and the resulting Voronoi geometry is more heterogeneous, the agreement remains strong, although the discrepancies are slightly more visible, particularly in the tail of the CDF and in the bin-probability comparison.

Table~\ref{tab:cdf_pdf_validation} summarizes the numerical CDF/PDF validation errors over the 100 repetitions. For Sim1, the mean KS error is 0.008182, with mean $\mathrm{CDF}\text{-}L^1$ and $\mathrm{CDF}\text{-}L^2$ errors equal to 0.001071 and $4\times 10^{-6}$, respectively. The mean PDF bin discrepancies are also small, with $\mathrm{PDF\_bin\_IAE}=0.048744$ and $\mathrm{PDF\_bin\_ISE}=9.6\times 10^{-5}$. Sim2 exhibits similarly strong agreement, with mean KS error 0.008529, mean $\mathrm{CDF}\text{-}L^1$ error 0.001512, and mean $\mathrm{PDF\_bin\_IAE}=0.046503$. Sim3 is the most challenging case, but the validation remains satisfactory: the mean KS error is 0.008985, the mean $\mathrm{CDF}\text{-}L^1$ error is 0.003929, and the mean $\mathrm{PDF\_bin\_IAE}$ is 0.055140. Overall, these results provide strong numerical support for the derived CDF and PDF formulas for $L_2$.

\begin{table}[htbp]
\centering
\caption{Distributional validation errors for the CDF and PDF of $L_2$ over 100 repetitions. For each scenario, the table reports the mean, maximum, standard deviation, median, and 95th percentile of the KS error, $\mathrm{CDF}\text{-}L^1$, $\mathrm{CDF}\text{-}L^2$, $\mathrm{PDF\_bin\_IAE}$, and $\mathrm{PDF\_bin\_ISE}$.}
\label{tab:cdf_pdf_validation}
\footnotesize{
\begin{tabular}{llccccc}
\hline
Scenario & Statistic & KS & CDF-$L^1$ & CDF-$L^2$ & PDF\_bin\_IAE & PDF\_bin\_ISE \\
\hline
Sim1 & Mean   & 0.008182 & 0.001071 & 0.000004 & 0.048744 & 0.000096 \\
Sim1 & Max    & 0.018182 & 0.002528 & 0.000021 & 0.063793 & 0.000202 \\
Sim1 & STD    & 0.002486 & 0.000378 & 0.000004 & 0.005661 & 0.000025 \\
Sim1 & Median & 0.007664 & 0.001017 & 0.000003 & 0.049088 & 0.000094 \\
Sim1 & Q95    & 0.012233 & 0.001740 & 0.000011 & 0.058129 & 0.000136 \\
\hline
Sim2 & Mean   & 0.008529 & 0.001512 & 0.000006 & 0.046503 & 0.000097 \\
Sim2 & Max    & 0.015163 & 0.003303 & 0.000021 & 0.062211 & 0.000208 \\
Sim2 & STD    & 0.002183 & 0.000503 & 0.000004 & 0.005845 & 0.000028 \\
Sim2 & Median & 0.008196 & 0.001409 & 0.000006 & 0.046525 & 0.000093 \\
Sim2 & Q95    & 0.011839 & 0.002559 & 0.000015 & 0.055956 & 0.000147 \\
\hline
Sim3 & Mean   & 0.008985 & 0.003929 & 0.000021 & 0.055140 & 0.000097 \\
Sim3 & Max    & 0.015427 & 0.007795 & 0.000073 & 0.071621 & 0.000152 \\
Sim3 & STD    & 0.002385 & 0.001441 & 0.000015 & 0.005846 & 0.000020 \\
Sim3 & Median & 0.008876 & 0.003732 & 0.000017 & 0.053882 & 0.000094 \\
Sim3 & Q95    & 0.013163 & 0.006530 & 0.000053 & 0.065319 & 0.000133 \\
\hline
\end{tabular}
}
\end{table}

\subsection{Moment validation}

We next retain the moment-validation experiment used in the previous version of the manuscript. For each repetition, we estimate the first four raw moments of $L_2$,
\[
\mathbb{E}(L_2),\quad \mathbb{E}(L_2^2),\quad \mathbb{E}(L_2^3),\quad \mathbb{E}(L_2^4),
\]
as well as the cosine-similarity moments
\[
\mathbb{E}[\cos^2(L_2)],\quad \mathbb{E}[\cos^4(L_2)],\quad \mathbb{E}[\cos^6(L_2)].
\]
The empirical estimates are then compared with the theoretical values derived from Equations~(21)--(23). Following the earlier setup, we summarize the accuracy using the absolute relative error (ARE), and report its mean, maximum, and standard deviation over the 100 repetitions.

Table~\ref{tab:moment_validation} reports the resulting moment-validation summary. The agreement remains strong across all three scenarios. In Sim1, the empirical estimates of the first four moments are very close to the corresponding theoretical values; for example, the theoretical value of $\mathbb{E}(L_2)$ is 0.175613, compared with a Monte Carlo mean of 0.175525, and the mean ARE is 0.003621. Similar agreement is observed for $\mathbb{E}(L_2^2)$, $\mathbb{E}(L_2^3)$, and $\mathbb{E}(L_2^4)$, as well as for the cosine moments. In Sim2, the mean AREs for the first four moments range from 0.004739 to 0.025302, while the cosine-moment AREs range from 0.000522 to 0.001307. Sim3 is the most challenging setting, due to the concentration of sites in two polar caps, but the overall agreement is still satisfactory: the mean AREs for the first four moments range from 0.004573 to 0.010899, while the cosine-moment AREs range from 0.003769 to 0.007629. These results are consistent with the conclusion of the original numerical study: the moment formulas derived in Section~4 are numerically accurate and stable across different site configurations.

Taken together, Tables~\ref{tab:cdf_pdf_validation} and \ref{tab:moment_validation} show that the theoretical results are accurate both at the level of the \emph{entire distribution} and at the level of \emph{selected moments}. The distributional table and Figure~\ref{fig:cdf_pdf_validation} demonstrate that the CDF and PDF derived in Section~3.3 agree closely with Monte Carlo simulations under all three site configurations, while the moment table confirms the correctness of the formulas in Section~4.

\begin{table}[htbp]
\centering
\caption{Validation of theoretical moments of $L_2$ and cosine similarity over 100 repetitions. For each scenario, the table reports the theoretical value (``Proposed Method''), the Monte Carlo mean (``Mean MC''), and the mean, maximum, and standard deviation of the absolute relative error (ARE).}
\label{tab:moment_validation}
\resizebox{\textwidth}{!}{
\begin{tabular}{llccccccc}
\hline
Scenario & Statistic & $\mathbb{E}(L_2)$ & $\mathbb{E}(L_2^2)$ & $\mathbb{E}(L_2^3)$ & $\mathbb{E}(L_2^4)$ & $\mathbb{E}[\cos^2(L_2)]$ & $\mathbb{E}[\cos^4(L_2)]$ & $\mathbb{E}[\cos^6(L_2)]$ \\
\hline
Sim1 & Proposed Method & 0.175613 & 0.039377 & 0.010446 & 0.003149 & 0.961656 & 0.926219 & 0.893363 \\
Sim1 & Mean MC         & 0.175525 & 0.039332 & 0.010427 & 0.003141 & 0.961698 & 0.926296 & 0.893469 \\
Sim1 & Mean ARE        & 0.003621 & 0.007000 & 0.011128 & 0.016246 & 0.000272 & 0.000521 & 0.000754 \\
Sim1 & Max ARE         & 0.012754 & 0.021975 & 0.036421 & 0.050402 & 0.000846 & 0.001604 & 0.002281 \\
Sim1 & STD ARE         & 0.002839 & 0.005385 & 0.008060 & 0.011470 & 0.000210 & 0.000408 & 0.000595 \\
\hline
Sim2 & Proposed Method & 0.194494 & 0.053283 & 0.018804 & 0.007977 & 0.949289 & 0.905322 & 0.866632 \\
Sim2 & Mean MC         & 0.194529 & 0.053294 & 0.018803 & 0.007972 & 0.949275 & 0.905292 & 0.866587 \\
Sim2 & Mean ARE        & 0.004739 & 0.010384 & 0.017194 & 0.025302 & 0.000522 & 0.000946 & 0.001307 \\
Sim2 & Max ARE         & 0.016917 & 0.041415 & 0.072559 & 0.107409 & 0.002042 & 0.003570 & 0.004744 \\
Sim2 & STD ARE         & 0.003496 & 0.007899 & 0.013498 & 0.019106 & 0.000398 & 0.000715 & 0.000967 \\
\hline
Sim3 & Proposed Method & 0.621943 & 0.506277 & 0.455513 & 0.433199 & 0.620328 & 0.462475 & 0.380827 \\
Sim3 & Mean MC         & 0.622239 & 0.506457 & 0.455587 & 0.433192 & 0.620147 & 0.462158 & 0.380424 \\
Sim3 & Mean ARE        & 0.004573 & 0.007129 & 0.009159 & 0.010899 & 0.003769 & 0.006062 & 0.007629 \\
Sim3 & Max ARE         & 0.012521 & 0.017927 & 0.024909 & 0.030060 & 0.009463 & 0.017126 & 0.023039 \\
Sim3 & STD ARE         & 0.003168 & 0.004867 & 0.006344 & 0.007574 & 0.002570 & 0.004154 & 0.005261 \\
\hline
\end{tabular}
}
\end{table}

\subsection{Computational efficiency}

We further evaluated the computational speed of the proposed method relative to two generic alternatives, namely direct numerical integration on the sphere and brute-force Monte Carlo simulation. The experiment was conducted under the same three sphere-level scenarios, Sim1--Sim3, using fixed seed configurations. To ensure a fair comparison, the two baseline methods were first calibrated so that their CDF and PDF outputs were sufficiently close to the proposed theoretical evaluator. Specifically, the proposed method was treated as the reference solution, and the resolutions of numerical integration and Monte Carlo were increased until they met prescribed discrepancy thresholds for the CDF and PDF. Runtime was then measured under these calibrated settings. For moment computation, we used the same calibrated settings determined by the CDF/PDF comparison rather than introducing a separate calibration criterion.

Table~\ref{tab:calibrated_settings} reports the calibrated resolutions. For the CDF, numerical integration reached the prescribed accuracy threshold with a $45\times 90$ grid in all three scenarios, while Monte Carlo required 20,000 samples in Sim1 and 10,000 samples in Sim2 and Sim3. For the PDF, the required resolutions were higher, as expected: numerical integration used a $90\times 180$ grid in Sim1 and Sim2 and a $120\times 240$ grid in Sim3, while Monte Carlo required 20,000 samples in Sim1 and Sim3 and 10,000 samples in Sim2.

\begin{figure}[htbp]
\centering

\begin{minipage}{0.32\textwidth}
    \centering
    \includegraphics[width=\textwidth]{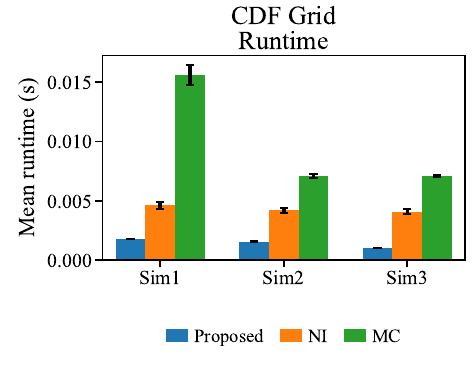}
\end{minipage}
\hfill
\begin{minipage}{0.32\textwidth}
    \centering
    \includegraphics[width=\textwidth]{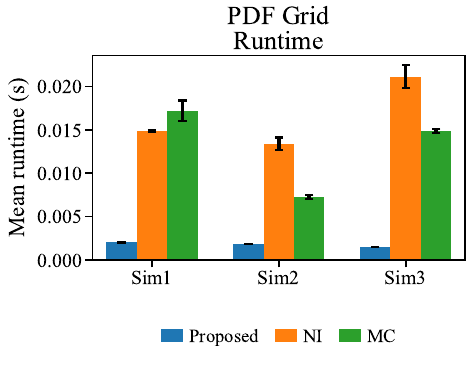}
\end{minipage}
\hfill
\begin{minipage}{0.32\textwidth}
    \centering
    \includegraphics[width=\textwidth]{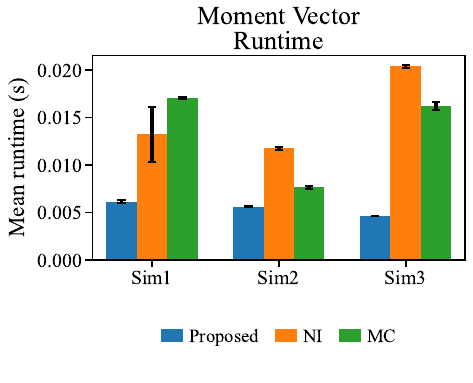}
\end{minipage}

\caption{Mean runtime comparison of the proposed method, numerical integration, and Monte Carlo under the calibrated settings. Error bars represent one standard deviation over repeated timing runs.}
\label{fig:runtime_bars}
\end{figure}

Under these calibrated settings, the proposed method is consistently the fastest for both CDF and PDF evaluation. As shown in Table~\ref{tab:runtime_speedup} and Figure~\ref{fig:runtime_bars}, for the CDF grid the proposed method is approximately 2.49--4.09 times faster than numerical integration and 4.67--8.40 times faster than Monte Carlo across the three scenarios. For the PDF grid, the advantage is even more pronounced: the proposed method is about 7.70--14.27 times faster than numerical integration and 3.97--10.05 times faster than Monte Carlo. Thus, for recovery of the full distribution of $L_2$, the optimized theoretical evaluator yields a substantial computational gain over both approximation-based alternatives.

\begin{figure}[htbp]
\centering

\begin{minipage}{0.32\textwidth}
    \centering
    \includegraphics[width=\textwidth]{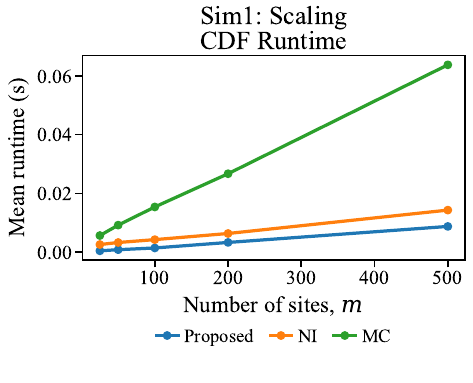}
\end{minipage}
\hfill
\begin{minipage}{0.32\textwidth}
    \centering
    \includegraphics[width=\textwidth]{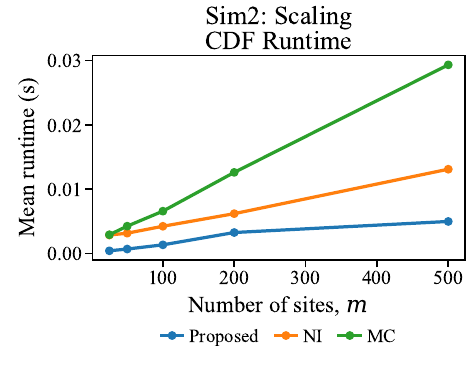}
\end{minipage}
\hfill
\begin{minipage}{0.32\textwidth}
    \centering
    \includegraphics[width=\textwidth]{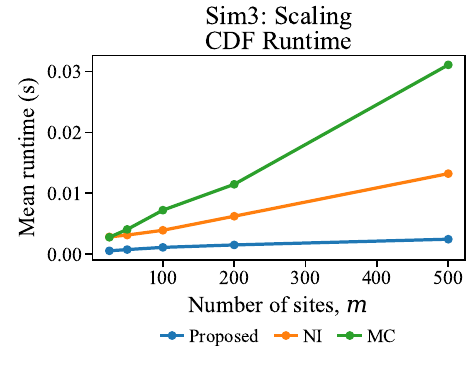}
\end{minipage}

\caption{Scaling study for CDF evaluation over $m\in\{25,50,100,200,500\}$. Across all three scenarios, the proposed method remains faster than numerical integration and Monte Carlo throughout the tested range.}
\label{fig:runtime_scaling}
\end{figure}

We also recorded the runtime for evaluating the moment vector under the same calibrated settings. Although moment computation was not used in the calibration step, the proposed method remained faster than both baselines in all three scenarios, with speedups of about 1.89--4.30 relative to numerical integration and 1.42--3.44 relative to Monte Carlo. Finally, Figure~\ref{fig:runtime_scaling} shows the scaling behavior for CDF evaluation over $m\in\{25,50,100,200,500\}$. In all three scenarios, the runtime of the proposed method increases with the number of sites, as expected, but remains below that of both competing approaches throughout the tested range. Overall, these results indicate that, once the repeated triangle-level evaluations are efficiently implemented, the proposed framework is not only theoretically rigorous but also computationally attractive for repeated evaluation of the distribution of $L_2$.

The computational advantage of the proposed method comes from simplifying
the numerical integration task. Instead of performing direct numerical
integration over the two-dimensional spherical surface, the proposed method
reduces the problem to repeated one-dimensional evaluations over arc-distance
thresholds after the Voronoi cells have been triangulated. Once the spherical
Voronoi diagram and triangle parameters are precomputed, the CDF, PDF, and
moment calculations are obtained by summing explicit triangle-level
contributions. This lower-dimensional deterministic structure explains why the
proposed method is faster than direct numerical integration and avoids the
large number of random query points required by Monte Carlo approximation.

\begin{table}[htbp]
\centering
\caption{Calibrated baseline settings for the computational-efficiency experiment. Numerical integration and Monte Carlo were first tuned so that their CDF and PDF outputs were sufficiently close to the proposed theoretical evaluator. The table reports the first resolution satisfying the prescribed discrepancy threshold in each scenario.}
\label{tab:calibrated_settings}
\begin{tabular}{llccc}
\hline
Scenario & Task & Method & Resolution & Error \\
\hline
Sim1 & CDF grid & Numerical Integration & $45\times 90$ & 0.006302 \\
Sim1 & CDF grid & Monte Carlo & 20000 & 0.008678 \\
Sim1 & PDF grid & Numerical Integration & $90\times 180$ & 0.043961 \\
Sim1 & PDF grid & Monte Carlo & 20000 & 0.045033 \\
\hline
Sim2 & CDF grid & Numerical Integration & $45\times 90$ & 0.004789 \\
Sim2 & CDF grid & Monte Carlo & 10000 & 0.007393 \\
Sim2 & PDF grid & Numerical Integration & $90\times 180$ & 0.042733 \\
Sim2 & PDF grid & Monte Carlo & 10000 & 0.052557 \\
\hline
Sim3 & CDF grid & Numerical Integration & $45\times 90$ & 0.006870 \\
Sim3 & CDF grid & Monte Carlo & 10000 & 0.008092 \\
Sim3 & PDF grid & Numerical Integration & $120\times 240$ & 0.032645 \\
Sim3 & PDF grid & Monte Carlo & 20000 & 0.045583 \\
\hline
\end{tabular}
\end{table}

\begin{table}[htbp]
\centering
\caption{Runtime comparison and speedup factors for the computational-efficiency experiment. The proposed method is compared with numerical integration (NI) and Monte Carlo (MC) under the calibrated settings in Table~\ref{tab:calibrated_settings}. Speedup is reported as baseline runtime divided by proposed runtime.}
\label{tab:runtime_speedup}
\resizebox{\textwidth}{!}{
\begin{tabular}{llcccccc}
\hline
Scenario & Task & Proposed (s) & NI (s) & MC (s) & NI / Proposed & MC / Proposed \\
\hline
Sim1 & CDF grid      & 0.001778 & 0.004432 & 0.014931 & 2.49 & 8.40 \\
Sim1 & PDF grid      & 0.002023 & 0.015578 & 0.014965 & 7.70 & 7.40 \\
Sim1 & Moment vector & 0.006298 & 0.011899 & 0.015984 & 1.89 & 2.54 \\
\hline
Sim2 & CDF grid      & 0.001537 & 0.004343 & 0.007172 & 2.83 & 4.67 \\
Sim2 & PDF grid      & 0.001834 & 0.014251 & 0.007275 & 7.77 & 3.97 \\
Sim2 & Moment vector & 0.005578 & 0.010719 & 0.007894 & 1.92 & 1.42 \\
\hline
Sim3 & CDF grid      & 0.000991 & 0.004053 & 0.007460 & 4.09 & 7.53 \\
Sim3 & PDF grid      & 0.001522 & 0.021723 & 0.015293 & 14.27 & 10.05 \\
Sim3 & Moment vector & 0.004679 & 0.020121 & 0.016107 & 4.30 & 3.44 \\
\hline
\end{tabular}
}
\end{table}

\section{Discussion}
\label{sec:conclusion}

In this paper, we studied the distributional properties of nearest-site angular
distances on the sphere. We first derive the CDF and PDF of $L_0(\triangle PBC)$, the
angular great-circle distance from a vertex $P$ of the spherical triangle
$\triangle PBC$ to a uniformly distributed random interior point $Q$. Using
polygon decomposition, we extend these results to cases in which the spherical
triangle is replaced by a convex spherical polygon, yielding $L_1(\Gamma,P)$.
Using spherical Voronoi diagrams, polygon triangulation, and numerical
integration, we obtain computable integral representations for arbitrary raw
moments of $L_2(\{P_1,\ldots,P_m\})$, the shortest angular great-circle
distance from a uniformly distributed random point $Q$ to a set of prespecified
seeds $\{P_1,\ldots,P_m\}$ on a sphere. Finally, we compute moments for
$L_2$ and $\cos(L_2)$, both of which are relevant to nearest-site analysis,
directional data analysis, spherical clustering, and cosine-similarity-based
machine learning methods.

Extensive Monte Carlo simulations were conducted to validate our theoretical results. Overall, the theoretical and the empirical results are highly consistent. As expected, higher order moments for both $L_{2}$ and $\cos L_{2}$ have larger AREs than the lower moments, because in general larger samples are required in order to obtain estimates of similar quality for higher order moments~\citep{casella2002statistical}. The computational comparison also illustrates the practical value of our approach: By reducing the global nearest-site distance problem to a finite collection of cap--triangle intersection calculations over triangulated spherical Voronoi cells, the method avoids relying exclusively on brute-force Monte Carlo simulation or generic surface integration.

Note that the assumption of convexity of a spherical polygon in deriving the CDF and PDF of $L_{1}$ is adopted for convenience rather than necessity. We believe the CDF/PDF of $L_{PQ}$, for $Q$ uniformly distributed in an arbitrary spherical triangle $\Delta$ of which $P \notin \Delta$, can be derived based on techniques we developed in this study. Since it is always possible to divide a non-convex spherical polygon into disjoint spherical triangles (not necessarily pivoting at a prespecified interior point $P$ if it is not convex), we can aggregate the CDF/PDF with exterior point $P$ to obtain the desired formula for a non-convex polygon. Furthermore, every smooth geometric object $O$ on $S^2$ can be approximated by spherical polygons, and hence by a disjoint union of spherical triangles $\Delta_{j}$. Therefore, the distribution of the arc distance between an arbitrary prespecified point $P \in S^{2}$ and $Q$ uniformly distributed on $O$ can be approximated by our method. How to achieve the best accuracy for this approximation with a fixed number of spherical triangles/polygons, and thereby attain the best trade-off between computational cost and accuracy, is an interesting research topic that warrants future study. 

Another natural direction for future work is to extend the present framework
from the two-dimensional sphere $S_R^2$ to higher-dimensional spheres or, more
generally, to higher-dimensional Riemannian manifolds. Such an extension is
conceptually appealing because nearest-site distance distributions and Voronoi
decompositions arise in many high-dimensional geometric and statistical
settings. However, the problem becomes substantially more difficult beyond
$S_R^2$. In higher dimensions, Voronoi cells are no longer spherical polygons
but higher-dimensional geodesic polytopes, and their decomposition requires
spherical or manifold simplices rather than triangles. The analogue of the
cap--triangle intersection area used in this paper becomes a cap--simplex
volume problem, whose boundary structure can involve many more geometric
cases and is generally much harder to express in closed or semi-closed form.
For general manifolds, additional complications arise from curvature variation,
possible lack of global geodesic convexity, and the absence of simple
trigonometric identities comparable to those available on the sphere. Developing
efficient and accurate methods for such higher-dimensional or manifold-valued
extensions is therefore an important but challenging topic for future research.

In summary, our study shows that nearest-site distance problems on spherical
domains can be analyzed without reducing the geometry to a planar
approximation or relying solely on Monte Carlo methods. The key idea is to
use the spherical Voronoi structure to localize the problem and then
aggregate triangle-level distributional contributions. This perspective
provides both a mathematical description of the underlying random distance
and a practical computational framework for evaluating its distributional
summaries. We expect that the same decomposition principle and other
mathematical techniques we developed in this study may be useful for other
stochastic distance problems involving spherical spatial data, directional
data, and non-Euclidean geometric domains.

\section*{Declaration of generative AI and AI-assisted technologies in the manuscript preparation process}

During the preparation of this work, the author(s) used ChatGPT for spelling and grammar checking. The author(s) reviewed and edited the output as needed and take full responsibility for the content of the published article.

%

\bibliography{sphericalVoronoiDistrMoments}


\appendix

\section{Definitions of useful geometric concepts of spherical triangles}\label{sec:append-def}

\begin{equation}
  \label{eq:SphrArcLen}
  D_{AB} = R\varphi_{AB}, \qquad L_{AB}=\varphi_{AB},
\end{equation}
where $L_{AB}$ denotes the angular great-circle distance in radians and
$D_{AB}$ denotes the corresponding physical arc length. This convention is
consistent with the notation used in the main text.

The area of the spherical cap, as shown in Fig. \ref{fig:SphrCap}, is

\begin{equation}
  \label{eq:SphrCapArea}
  A_{cap} = 2 \pi R^2 (1- \cos \theta )
\end{equation}

The area of the spherical sector, as shown in Fig. \ref{fig:Sphrfan}, is

\begin{equation}
  \label{eq:SphrSctArea}
  A_{fan} = \alpha R^2 (1- \cos \theta )
\end{equation}

\section{Moments of Cosine Similarity}\label{sec:append-cosine-similarity}

Using the triangle-level density of $L_0$ in Corollary~\ref{cor:PDF_L0}, we first
derive selected moments of the cosine similarity $\cos(L_0)$. The corresponding
moments for $L_1$ and $L_2$ can then be obtained by the same area-weighted
aggregation used in Section~\ref{sec:Moments}.
\begin{equation}
  \begin{split}
    \label{eq:ECos2-initial}
    E[\cos^2(L_0)] &= \int^{ \infty }_{-\infty} \cos^2(r) f_{L_0}(r) dr \\
    &=\frac{R^2}{\Omega}( \int_0^{ h } \cos^2(r) \cdot \alpha_P \sin r dr\\
    &\quad + \int_{h}^{  L_{PB} } \cos^2(r)\cdot \{ \alpha_1 \sin r + \eta [\alpha'_r \cos r+\beta'_r+( \alpha_2-\alpha_r) \sin r ] \} dr\\
    &\quad +\int_{L_{PB}}^{  L_{PC} } \cos^2(r)\cdot [\beta'_r- \omega'_r \cos r + \omega_r \sin r ] dr ) \\
    &= \frac{R^2}{\Omega}( \mathrm{Term}_1 + \mathrm{Term}_2 + \mathrm{Term}_3 ).
  \end{split}
\end{equation}

The first two terms in the above equation are derived as follows.
\begin{equation}
  \label{eq:Term1}
  \begin{split}
    \mathrm{Term}_1 &= \int_0^{ h } \cos^2(r) \cdot \alpha_P \sin r dr = \frac{-1}{3} \alpha_P \cdot \cos^3(r)|_0^{ h } \\
    &= \frac{-1}{3} \alpha_P \cdot (\cos^3(h) - 1) = \frac{1}{3} \alpha_P (1 - \cos^3(h) ).
  \end{split}
\end{equation}

\begin{equation}
  \label{eq:Term2}
  \begin{split}
    \mathrm{Term}_2 &= \int_{h}^{  L_{PB} } \cos^2(r)\cdot \{ \alpha_1 \sin r + \eta [\alpha'_r \cos r+\beta'_r+( \alpha_2-\alpha_r) \sin r ]\} dr \\
    &= \int_{h}^{  L_{PB} } \cos^2(r)\cdot \alpha_1 \sin r dr +  \eta I_{\eta} \\
    &=  \frac{1}{3}\alpha_1 [\cos^3 ( h ) - \cos^3( L_{PB})] +\eta I_{\eta}.
  \end{split}
\end{equation}
Here
\begin{equation}
  \label{eq:I_eta}
  \begin{split}
    I_{\eta}&= \int_{h}^{  L_{PB} } \cos^2(r)[\alpha'_r \cos r+\beta'_r+( \alpha_2-\alpha_r) \sin r ] dr\\
    &= \int_{h}^{  L_{PB} }(\cos^3r \alpha'_r - \cos^2r \sin r \alpha_r) dr +\int_{h}^{  L_{PB} }\cos^2r \beta'_r dr \\
    &\quad +\int_{h}^{  L_{PB} } \alpha_2 \cos^2r \sin r dr = I_{\eta,1} + I_{\eta,2}+ I_{\eta,3},
  \end{split}
\end{equation}
where 
\begin{equation}
  \label{eq:I_eta1}
  \small
  \begin{split}
    I_{\eta,1} &= \int_{h}^{  L_{PB} }(\cos^3r \alpha'_r - \cos^2r \sin r \alpha_r) dr \\
    &= \frac{1}{3} \alpha_r \cos^3r|^{L_{PB}}_{h} + \frac{2}{3} \int_{h}^{  L_{PB} } \cos^3r \alpha'_r dr \\
    &= \frac{1}{3} [\arccos \frac{\tan h}{\tan L_{PB}} \cos^3 L_{PB} - 0 ]
    + \frac{2}{3} \int_{h}^{  L_{PB} } \frac{\tan h \cos^2 r}{\sin r \sqrt{\tan^2r - \tan^2h}}  dr \\
    &= \frac{1}{3}\arccos \frac{\tan h}{\tan L_{PB}} \cos^3 L_{PB} + \frac{2}{3} \cdot I_{\eta,1p}.
  \end{split}
\end{equation}
\begin{equation}
  \label{eq:I_eta1p}
  \small
  \begin{split}
    I_{\eta,1p} &= \int_{h}^{  L_{PB} } \frac{\tan h \cos^2 r}{\sin r \sqrt{\tan^2r - \tan^2h}}  dr\\
    &= \tan h \Bigr( \int_{h}^{  L_{PB} } \frac{1}{\sin r \sqrt{\tan^2r - \tan^2h}}  dr
    - \int_{h}^{  L_{PB} } \frac{ \sin r}{ \sqrt{\tan^2r - \tan^2h}}  dr \Bigl) \\
    &= -\frac{1}{2} \arctan \frac{2 \tan h  \sqrt{\tan^2 r-\tan^2 h}\sqrt{\tan^2 r+1}}{ \tan^2 r-\tan^2 r \tan^2 h-2 \tan^2 h } \Big|_{h}^{  L_{PB} }\\
    &\quad -\frac{ \sqrt{\tan^2 r - \tan^2 h}}{ \tan h (1 +\tan^2 h ) \sqrt{1 +\tan^2 r}} \Big|_{h}^{  L_{PB} }\\
    &=  -\frac{ 1}{2} \arctan \frac{2 \tan h  \sqrt{\tan^2 {  L_{PB} }-\tan^2 h}\sqrt{\tan^2 {  L_{PB} }+1}}{ \tan^2 {  L_{PB} }-\tan^2 {  L_{PB} } \tan^2 h-2 \tan^2 h } \\
    &\quad -\frac{ \tan h \sqrt{\tan^2 {  L_{PB} } - \tan^2 h}}{(1 +\tan^2 h ) \sqrt{1+\tan^2 {  L_{PB} }}}.
  \end{split}
\end{equation}

Note that the following two technical results are needed in deriving the above integral. Let $u = \tan r$, $a= \tan h$, we have
\begin{equation}
  \label{eq:eta_1pDetail01}
  \small
  \begin{split}
    &\int \frac{1}{\sin r \sqrt{\tan^2r - \tan^2h}} dr = \int \frac{1}{u \sqrt{u^2 - a^2}  \sqrt{1+ u^2}} du  \\
    &=  \frac{ \sqrt{-a^2} }{2 a^2} \Bigg[\ln \frac{ \sqrt{u^2 + 1} - 1}{ \sqrt{-a^2} - \sqrt{u^2 - a^2}} -\ln( \sqrt{-a^2}) \\
    &\quad +\frac{ \sqrt{u^2 + 1} - 1}{ \sqrt{-a^2} - \sqrt{u^2 - a^2} }
    - \frac{ a^2 (\sqrt{u^2 + 1} - 1)}{ \sqrt{-a^2} - \sqrt{u^2 - a^2}}
    + \frac{ \sqrt{-a^2} ( \sqrt{u^2 + 1} - 1)^2}{ (\sqrt{-a^2} - \sqrt{u^2 - a^2} )^2}   \Bigg] \\
    &= \frac{ i }{2a}\ln \frac{u^2-u^2 a^2-2 a^2 + i \cdot 2 a  \sqrt{u^2-a^2}\sqrt{u^2+1}}{u^2 (a^2+1)^2} \\
    &= -\frac{ 1}{2a} \arctan \frac{2 a  \sqrt{u^2-a^2}\sqrt{u^2+1}}{ u^2-u^2 a^2-2 a^2 } +\mathrm{Const},\\
    \mathrm{Const} &:=
    \begin{cases}
      -\frac{i}{2a} \ln{(a^2+1)}, & u^2-u^2 a^2-2 a^2 > 0, \\
      -\frac{i}{2a} \ln{(a^2+1)}-\frac{ \pi}{2a},~ & u^2-u^2 a^2-2 a^2 < 0.
    \end{cases}
  \end{split}
\end{equation}
Note that only the first term of the results in Equation~\eqref{eq:eta_1pDetail01} depends on variable $u := \tan r$, so the choice of $\mathrm{Const}$ is irrelevant in computing the corresponding definite integral used in Equation~\eqref{eq:I_eta1p}.

Again let $u = \tan^2 r$, $b = \tan^2 h$, we have
\begin{equation}
  \label{eq:eta_1pDetail02}
  \begin{split}
    &\quad \int \frac{ \sin r}{ \sqrt{\tan^2r - \tan^2h}} dr = \int \frac{ 1}{ 2 \sqrt{u - b} (1+u)^{3/2}} du  \\
    &= \frac{ \sqrt{u - b}}{(b + 1) \sqrt{u + 1}} = \frac{ \sqrt{\tan^2 r - \tan^2 h}}{(1 +\tan^2 h ) \sqrt{1+\tan^2 r}}.
  \end{split}
\end{equation}

The rest two terms, $I_{\eta,2}, I_{\eta,3}$ are provided as follows.
\begin{equation}
  \label{eq:eta_2}
  \begin{split}
    I_{\eta,2} &= \int_{h}^{  L_{PB} }\cos^2r \beta'_r dr = -\int_{h}^{  L_{PB} } \frac{\sin h \cos^3 r}{\sin r \sqrt{\sin^2r - \sin^2 h}} dr \\
    &= - ( \arctan \frac{ \sqrt{\sin^2r - \sin^2 h}}{\sin h} - \sin h \sqrt{\sin^2r - \sin^2 h} ) \Big|_{h}^{  L_{PB} }\\
    &= \sin h \sqrt{\sin^2 L_{PB} - \sin^2 h } - \arctan \frac{ \sqrt{\sin^2L_{PB} - \sin^2h }}{\sin h }.
  \end{split}
\end{equation}
\begin{equation}
  \label{eq:eta_3}
  \begin{split}
    I_{\eta,3} &= \int_{h}^{  L_{PB} } \alpha_2 \cos^2r \sin r dr = -\frac{1}{3} \alpha_2 \cos^3r \Big|^{L_{PB}}_{h} \\
    &= \frac{1}{3} \alpha_2 ( \cos^3 h - \cos^3 L_{PB} ).
  \end{split}
\end{equation}

In summary, we have
\begin{equation}
  \label{eq:I_eta_final}
  \small
  \begin{split}
    I_{\eta} &= I_{\eta,1} + I_{\eta,2} +I_{\eta,3} \\
    &= \frac{1}{3}\arccos \frac{\tan h}{\tan L_{PB}} \cos^3 L_{PB} -\frac{ 1}{3} \arctan \frac{2 \tan h  \sqrt{\tan^2 {  L_{PB} }-\tan^2 h}\sqrt{\tan^2 {  L_{PB} }+1}}{ \tan^2 {  L_{PB} }-\tan^2 {  L_{PB} } \tan^2 h-2 \tan^2 h } \\
    &\quad -\frac{ 2\tan h \sqrt{\tan^2 {  L_{PB} } - \tan^2 h}}{3(1 +\tan^2 h ) \sqrt{1+\tan^2 {  L_{PB} }}} +\sin h \sqrt{\sin^2 L_{PB} - \sin^2 h } \\
    &\quad -\arctan \frac{ \sqrt{\sin^2L_{PB} - \sin^2h }}{\sin h } +\frac{1}{3} \alpha_2 ( \cos^3 h - \cos^3 L_{PB} ).
  \end{split}
\end{equation}

$\mathrm{Term}_3$ is derived as follows.
\begin{equation}
  \label{eq:Term3}
  \small
  \begin{split}
    \mathrm{Term}_3 &= \int_{L_{PB}}^{  L_{PC} } \cos^2(r)\cdot [\beta'_r- \omega'_r \cos r + \omega_r \sin r ] dr \\
    &= -\int_{L_{PB}}^{  L_{PC} } \frac{\sin h \cos^3 r}{\sin r \sqrt{\sin^2 r - \sin^2 h} } dr  -\frac{1}{3} \omega_r \cos^3 r \Big|_{L_{PB}}^{L_{PC}}
    - \frac{2}{3} \int_{L_{PB}}^{L_{PC}} \omega '_r \cos^3 r dr\\
    &=  \sin h \sqrt{\sin^2 L_{PC} - \sin^2 h } - \arctan \frac{ \sqrt{\sin^2 L_{PC} - \sin^2 h }}{\sin h } \\
    &\quad -\sin h \sqrt{\sin^2 L_{PB} - \sin^2 h } + \arctan \frac{ \sqrt{\sin^2 L_{PB} - \sin^2 h }}{\sin h } \\
    &\quad + \frac{1}{3} \left[\cos^3 L_{PB} \arccos \frac{\tan h}{ \tan L_{PB}} - \cos^3 L_{PC} \arccos \frac{\tan h}{ \tan L_{PC}} \right] \\
    &\quad + \frac{2}{3} \int_{L_{PB}}^{L_{PC}} \cos^3 r \alpha '_r  dr\\
    &= \sin h \sqrt{\sin^2 L_{PC} - \sin^2 h } - \arctan \frac{ \sqrt{\sin^2 L_{PC} - \sin^2 h }}{\sin h } \\
    &\quad -\sin h \sqrt{\sin^2 L_{PB} - \sin^2 h } + \arctan \frac{ \sqrt{\sin^2 L_{PB} - \sin^2 h }}{\sin h } \\
    &\quad + \frac{1}{3}\cos^3 L_{PB} \arccos \frac{\tan h}{ \tan L_{PB}} - \frac{1}{3}\cos^3 L_{PC} \arccos \frac{\tan h}{ \tan L_{PC}} +\frac{2}{3} \mathrm{Term}_{4}.
  \end{split}
\end{equation}
Using Equation (\ref{eq:I_eta1p}), we have
\begin{equation}
  \label{eq:Term4}
  \begin{split}
    &\quad \mathrm{Term}_{4} = \int_{L_{PB}}^{L_{PC}} \cos^3 r \alpha '_r  dr = \int_{L_{PB}}^{L_{PC}} \frac{\tan h \cos^2 r}{\sin r \sqrt{\tan^2r - \tan^2h}}  dr \\
    &= \frac{ 1}{2 } \arctan \frac{2 \tan h  \sqrt{\tan^2 L_{PB}-\tan^2 h}\sqrt{\tan^2 L_{PB}+1}}{ \tan^2 L_{PB}-\tan^2 L_{PB} \tan^2 h-2 \tan^2 h } \\
    &\quad -\frac{ 1}{2} \arctan \frac{2 \tan h  \sqrt{\tan^2 L_{PC}-\tan^2 h}\sqrt{\tan^2 L_{PC}+1}}{ \tan^2 L_{PC}-\tan^2 L_{PC} \tan^2 h-2 \tan^2 h } \\
    &\quad -\frac{ \tan h \sqrt{\tan^2 L_{PC} - \tan^2 h}}{(1 +\tan^2 h ) \sqrt{1+\tan^2 L_{PC}}}
    +\frac{ \tan h \sqrt{\tan^2 L_{PB} - \tan^2 h}}{(1 +\tan^2 h ) \sqrt{1+\tan^2 L_{PB}}}.
  \end{split}
\end{equation}

In summary, we have 
\begin{equation}
  \label{eq:Ecos2}
  \begin{split}
    E[\cos^2(L_0)]
    &= \frac{R^2}{\Omega}\left(\mathrm{Term}_1+\mathrm{Term}_2+\mathrm{Term}_3\right),
  \end{split}
\end{equation}
where $\mathrm{Term}_1$, $\mathrm{Term}_2$, and $\mathrm{Term}_3$ are defined in
Equations~\eqref{eq:Term1}--\eqref{eq:Term3}. This compact representation is
used to avoid ambiguity in the normalization factor; any fully expanded form
must be multiplied by the same factor $R^2/\Omega$.

Similarly, the fourth and sixth cosine moments of $L_0$ can be written as follows.
\begin{equation}\label{eq:Ecos4}
  \scriptsize
  \begin{split}
    E\left[\cos^4(L_{0})\right] &= \int_{-\infty}^{+\infty} \cos^4  r \cdot f_{L_0}\left( r \right) dr \\
    &= \frac{R^2}{\Omega} \bigg(\int_0^h \cos^4  r \cdot \alpha_P \sin r dr \\
    &\quad +\int_h^{L_{PB}}\cos^4 r \cdot \left\{ \alpha_1 \sin r + \eta \left[ \alpha'_r \cos r + \beta'_r +\left( \alpha_2 - \alpha_r \right) \sin r \right] \right\} dr \\
    &\quad +\int_{L_{PB}}^{L_{PC}}\cos^4 r \cdot \left[ \beta'_r - \omega'_r \cos r + \omega_r \sin r \right] dr\bigg) \\
    &= \frac{R^2}{\Omega}\Bigg(\frac{1}{5}\left( \angle{MPC} \cos^5 \left( L_{PC} \right)  -\left(\angle {MPC}+ \alpha_2 \eta +\alpha_1 \right) \cos^5 \left( L_{PB} \right)  +\left(\alpha_2 \eta - \alpha_P + \alpha_1\right) \cos^5 h +\alpha_P \right) \\
    &\quad +\frac{ 4\eta -6}{5} \arctan\left( \frac{\sqrt {\cos^2 h -\cos^2 \left( L_{PB} \right)}} {\sin h}\right)    + \frac{6}{5}\arctan\left( \frac{\sqrt {\cos^2 h -\cos^2 \left( L_{PC} \right)}} {\sin h}\right) \\
    &\quad +\frac{ 4\eta -6  }{15} \sin h \left( \sin^2 \left( L_{PB} \right)-2 \cos^2 h -4 \right)\sqrt{\cos^2 h -\cos^2 \left( L_{PB} \right) }\\
    &\quad +\frac{2}{5}\sin h \left( \sin^2 \left( L_{PC} \right)-2 \cos^2 h -4 \right)\sqrt{ \cos^2 h -\cos^2 \left( L_{PC} \right)}\\
    &\quad +\left(\eta-1\right) \arcsin \left( \frac{\sin h}{\left| \sin \left( L_{PB} \right) \right|} \right)   +\arcsin \left( \frac{\sin h}{\left| \sin \left( L_{PC} \right) \right|} \right)   -  \eta \arcsin \left( \frac{\sin h}{\left| \sin h \right|} \right) \\
    &\quad -\frac{1}{5}\left(\cos^5 \left( L_{PC} \right) \arccos\left( \frac{\tan h}{\tan \left( L_{PC} \right)}\right) - \left(1+\eta\right) \cos^5 \left( L_{PB} \right) \arccos\left( \frac{\tan h }{\tan \left( L_{PB} \right)}\right) \right) \\
    &\quad -\frac{ \eta-1}{3}   \sin h \sqrt {\left(\sin \left( L_{PB} \right) -\sin h \right)\left(  \sin \left( L_{PB} \right) + \sin h  \right)}   \left( \sin^2 \left( L_{PB} \right) +2 \sin^2 h -6 \right)     \\
    &\quad -\frac{1}{3} \sin h \sqrt {\left(\sin \left( L_{PC} \right) -\sin h \right)\left(  \sin \left( L_{PC} \right) + \sin h  \right)}   \left( \sin^2 \left( L_{PC} \right) +2 \sin^2 h -6 \right) \\
    &\quad -\frac{2}{5}\left[
      \left(\angle{MPC}-\arccos\left(\frac{\tan h}{\tan L_{PC}}\right)\right)\cos^5 L_{PC}
      -\left(\angle{MPC}-\arccos\left(\frac{\tan h}{\tan L_{PB}}\right)\right)\cos^5 L_{PB}
    \right] \\
    &\quad +\frac{8}{5}\int_{L_{PB}}^{L_{PC}}\cos^5 r\,\alpha'_r\,dr
    \Bigg).
  \end{split}
\end{equation}

  \begin{equation}\label{eq:Ecos6}
    \scriptsize
    \begin{split}
      E\left[\cos^6(L_{0})\right] &= \int_{-\infty}^{+\infty} \cos^6  r \cdot f_{L_0}\left( r \right) dr \\
      &= \frac{R^2}{\Omega} \bigg(\int_0^h \cos^6  r \cdot \alpha_P \sin r      dr \\
      &\quad + \int_h^{L_{PB}}\cos^6 r \cdot \left\{ \alpha_1 \sin r + \eta \left[ \alpha'_r \cos r + \beta'_r +\left( \alpha_2 - \alpha_r \right) \sin r \right] \right\} dr \\
      &\quad + \int_{L_{PB}}^{L_{PC}}\cos^6 r \cdot \left[ \beta'_r - \omega'_r \cos r + \omega_r \sin r \right] dr \bigg) \\
      &=
      \dfrac{R^2}{\Omega} \Bigg(\dfrac{1}{7}\left( \cos^7 h  \left( \alpha_1 \eta - \alpha_P + \alpha_1 \right)-\cos^7 \left( L_{PB}\right) \left( \alpha_2 \eta - \angle{MPC} + \alpha_1 \right)+\angle{MPC} \cos^7\left( L_{PC} \right)+\alpha_P \right) \\
      &\quad +\dfrac{ 6\eta -8}{7} \arctan\left( \frac{\sqrt {\cos^2 h -\cos^2 \left( L_{PB} \right)}} {\sin h}\right)    + \frac{8}{7}\arctan\left( \frac{\sqrt {\cos^2 h -\cos^2 \left( L_{PC} \right)}} {\sin h}\right) \\
      &\quad -\dfrac{ 6\eta -8  }{105} \sin h \sqrt{\cos^2 h -\cos^2 \left( L_{PB} \right) }\left( \sin^2\left(L_{PB}\right)\left( 3 \sin^2\left(L_{PB}\right)-4\cos^2h-11\right)+8\cos^4 h +14\cos^2 h+23\right) \\
      &\quad -\dfrac{ 8  }{105} \sin h \sqrt{\cos^2 h -\cos^2 \left( L_{PC} \right) }\left( \sin^2\left(L_{PC}\right)\left( 3 \sin^2\left(L_{PC}\right)-4\cos^2h-11\right)+8\cos^4 h +14\cos^2 h+23\right)\\
      &\quad +\left(\eta-1\right) \arcsin \left( \frac{\sin h}{\left| \sin \left( L_{PB} \right) \right|} \right)   +\arcsin \left( \frac{\sin h}{\left| \sin \left( L_{PC} \right) \right|} \right)   -  \eta \arcsin \left( \frac{\sin h}{\left| \sin h \right|} \right) \\
      &\quad +\dfrac{1}{7}\left( \left(1+\eta\right) \cos^7 \left( L_{PB} \right) \arccos\left( \frac{\tan h }{\tan \left( L_{PB} \right)}\right)  -\cos^7 \left( L_{PC} \right) \arccos\left( \frac{\tan h}{\tan \left( L_{PC} \right)}\right)  \right) \\
      &\quad +\dfrac{ \eta-1}{15} \sin h \sqrt {\left(\sin^2 \left( L_{PB} \right) -\sin^2 h \right)}\left( 3 \sin^4\left( L_{PB}\right) + \left( 4 \sin^2 h -15\right) \sin^2 \left( L_{PB}\right) + 8 \sin^4 h -30 \sin^2 h +45\right) \\
      &\quad +\dfrac{ 1}{15} \sin h \sqrt {\left(\sin^2 \left( L_{PC} \right) -\sin^2 h \right)}\left( 3 \sin^4\left( L_{PC}\right) + \left( 4 \sin^2 h -15\right) \sin^2 \left( L_{PC}\right) + 8 \sin^4 h -30 \sin^2 h +45\right) \\
      &\quad -\frac{2}{7}\left[
        \left(\angle{MPC}-\arccos\left(\frac{\tan h}{\tan L_{PC}}\right)\right)\cos^7 L_{PC}
        -\left(\angle{MPC}-\arccos\left(\frac{\tan h}{\tan L_{PB}}\right)\right)\cos^7 L_{PB}
      \right] \\
      &\quad +\frac{12}{7}\int_{L_{PB}}^{L_{PC}}\cos^7 r\,\alpha'_r\,dr
      \Bigg).
    \end{split}
  \end{equation}

\end{document}